%% file: medima-template.tex
\documentclass[times,twocolumn,final]{elsarticle}
\usepackage{medima}
\usepackage{svg}
\usepackage{framed,multirow}

\usepackage{amssymb}
\usepackage{latexsym}
\usepackage{multirow}
\usepackage{amsmath,amssymb,amsfonts}
\usepackage{graphicx}    % For including images
\usepackage{subfig}
\usepackage{subcaption}
\usepackage{relsize}\usepackage{subdepth}\allowdisplaybreaks

\usepackage{url}
\usepackage{xcolor}
\usepackage[normalem]{ulem}
\usepackage{tabularray}
\usepackage{hyperref}

\definecolor{newcolor}{rgb}{.8,.349,.1}

\journal{Medical Image Analysis}

\begin{document}

\verso{Given-name Surname \textit{et~al.}}

\begin{frontmatter}

\title{Push the Boundary of SAM: A Pseudo-label Correction Framework for Medical Segmentation}%
% \tnotetext[tnote1]{This is an example for title footnote coding.}

\author[1]{Ziyi \snm{Huang} \fnref{fn1}}

\author[2,3]{Hongshan \snm{Liu}\fnref{fn1}}

\fntext[fn1]{Equally contributed first authors}

\author[4]{Haofeng \snm{Zhang}}
%% Third author's email
% \ead{author3@author.com}

\author[2,3]{Xueshen \snm{Li}}

\author[2,3]{Haozhe \snm{Liu}}

\author[5]{Fuyong \snm{Xing}}

\author[6]{Andrew F.  \snm{Laine}}

\author[6,7,8]{Elsa D.  \snm{Angelini}}

\author[1]{Christine P.  \snm{Hendon} \corref{cor1}}

\author[2,3]{Yu \snm{Gan} \corref{cor1}}

\cortext[cor1]{Corresponding author: 
cpf2115@columbia.edu, Department of Electrical Engineering, Columbia University, New York, NY 10027 USA (C.P. Hendon). ygan5@stevens.edu, Department of Biomedical Engineering, Stevens Institute of Technology, Hoboken, NJ 07030 USA (Y. Gan).}

% \nonumnote{This paper is supported in part by NIH-5R01HL14936 (CPH),  Cheung-Kong Innovation Doctoral Fellowship (ZH), NSF-2222739(YG), NSF-2239810 (YG), New Jersey Health Foundation (YG), USDA-2022-67021-36866 (YG)}

\nonumnote{The code is available at https://github.com/YGanLab/Pseudo-label-Correction-Framework}

\address[1]{Department of Electrical Engineering, Columbia University, New York, NY 10027 USA}
\address[2]{Department of Biomedical Engineering, Stevens Institute of Technology, Hoboken, NJ 07030 USA}
\address[3]{Center for Health Innovation, Stevens Institute of Technology, Hoboken, NJ 07030 USA}
\address[4]{Department of Industrial Engineering and Operations Research, Columbia University, New York, NY 10027 USA}
\address[5]{Department of Biostatistics and Informatics, Colorado School of Public Health, University of Colorado Anschutz Medical Campus, Aurora CO 80045, USA}
\address[6]{Department of Biomedical Engineering, Columbia University, New York, NY 10027 USA}
\address[7]{NIHR Imperial Biomedical Research Centre and ITMAT Data Science Group, Imperial College London, SW7 2BX London, U.K.}
\address[8]{Telecom Paris, LTCI, Institut Polytechnique de Paris, 91120 Palaiseau, France}

\received{20 Dec 2023}
\finalform{XXX}
\accepted{YYY}
\availableonline{ZZZ}
\communicated{AAA}

\begin{abstract}
%%%
Segment anything model (SAM) has emerged as the leading approach for zero-shot learning in segmentation tasks, offering the advantage of avoiding pixel-wise annotations. It is particularly appealing in medical image segmentation, where the annotation process is laborious and expertise-demanding. However, the direct application of SAM often yields inferior results compared to conventional fully supervised segmentation networks. An alternative approach is to use SAM as the initial stage to generate pseudo labels for further network training. However, the performance is limited by the quality of pseudo labels. In this paper, we propose a novel label correction framework to push the boundary of SAM-based segmentation. Our model utilizes a label quality evaluation module to distinguish between noisy labels and clean labels. This enables the correction of the noisy labels using an uncertainty-based self-correction module, thereby enriching the clean training set. Finally, we retrain the segmentation network with updated labels to optimize its weights for future predictions. One key advantage of our model is its ability to train deep networks using SAM-generated pseudo labels without relying on a set of expert-level annotations while attaining good segmentation performance.  We demonstrate the effectiveness of our proposed model on three public datasets, indicating its ability to improve segmentation accuracy and outperform baseline methods in label correction.
%%%%
\end{abstract}


\end{frontmatter}

\input{sections/introduction}

\input{sections/relatedwork}

\input{sections/method}

\input{sections/results}

\input{sections/Discussion}

\input{sections/conclusion}

%\linenumbers


\section*{CRediT authorship contribution statement}
\textbf{ZH:} Conceptualization, Methodology, Software, Validation, Formal analysis, Investigation, Data Curation, Visualization, Writing - Original Draft.
\textbf{HS Liu:} Conceptualization, Methodology, Software, Validation, Formal analysis, Investigation, Data Curation, Visualization, Writing - Original Draft.
\textbf{HZ:} Conceptualization, Methodology, Software.
\textbf{XL:} Software, Validation, Formal analysis, Visualization, Investigation.
\textbf{HZ Liu:} Software, Validation, Formal analysis, Visualization,Investigation, Data Curation.
\textbf{FX:} Conceptualization, Methodology, Investigation, Writing - Review \& Editing.
\textbf{AFL:} Conceptualization, Methodology, Investigation, Writing - Review \& Editing.
\textbf{EDA:} Conceptualization, Methodology, Investigation, Writing - Review \& Editing.
\textbf{CPH:} Conceptualization, Formal analysis, Funding acquisition
Investigation, Methodology, Project administration, Software, Supervision, Validation, Visualization Writing – original draft, Writing – review \& editing.
\textbf{YG:} Conceptualization, Formal analysis, Funding acquisition
Investigation, Methodology, Project administration, Software, Supervision, Validation, Visualization Writing – original draft, Writing – review \& editing.

% For additional terms see this website: Project administration, Funding acquisition.
%https://www.elsevier.com/researcher/author/policies-and-guidelines/credit-author-statement

\section*{Declaration of competing interest}
The authors declare that they have no conflict of interest.

\section*{Acknowledgments}
Funding: This paper is supported in part by NIH-5R01HL14936 (CPH),  Cheung-Kong Innovation Doctoral Fellowship (ZH), Stevens Provost Fellowship (HZ Liu), NSF-2222739(YG), NSF-2239810 (YG), New Jersey Health Foundation (YG), USDA/NIFA-2022-67021-36866 (YG).

% \section*{References}

% Please ensure that every reference cited in the text is also present in
% the reference list (and vice versa).

%%Harvard
\bibliographystyle{model2-names.bst}\biboptions{authoryear}
\bibliography{refs}

\end{document}

%% file: sections/introduction.tex
\section{Introduction}
%Deep learning has added a huge boost in the rapidly developing field of medical image analysis. %%%In clinical practice, a high performance segmentation method for organs and other tissue structures could greatly enhance the precision of medical diagnosis and computer guided therapy.
%However, 

The recent success of the segment anything model (SAM) \citep{kirillov2023segment} in the field of image segmentation sparks numerous discussions in computer vision \citep{zhang2023personalize,wang2023scaling,mo2023av,ji2023sam}. Segment anything model, with a zero-shot setting, was trained on diverse data, including over a billion masks. One of its significant advantages lies in its ability to segment without pixel-wise manual annotation, which is particularly appealing in medical image segmentation, considering the laborious and expertise-demanding nature of annotation. 

\begin{figure}[t]
\begin{center}
\includegraphics[width=0.5\textwidth]{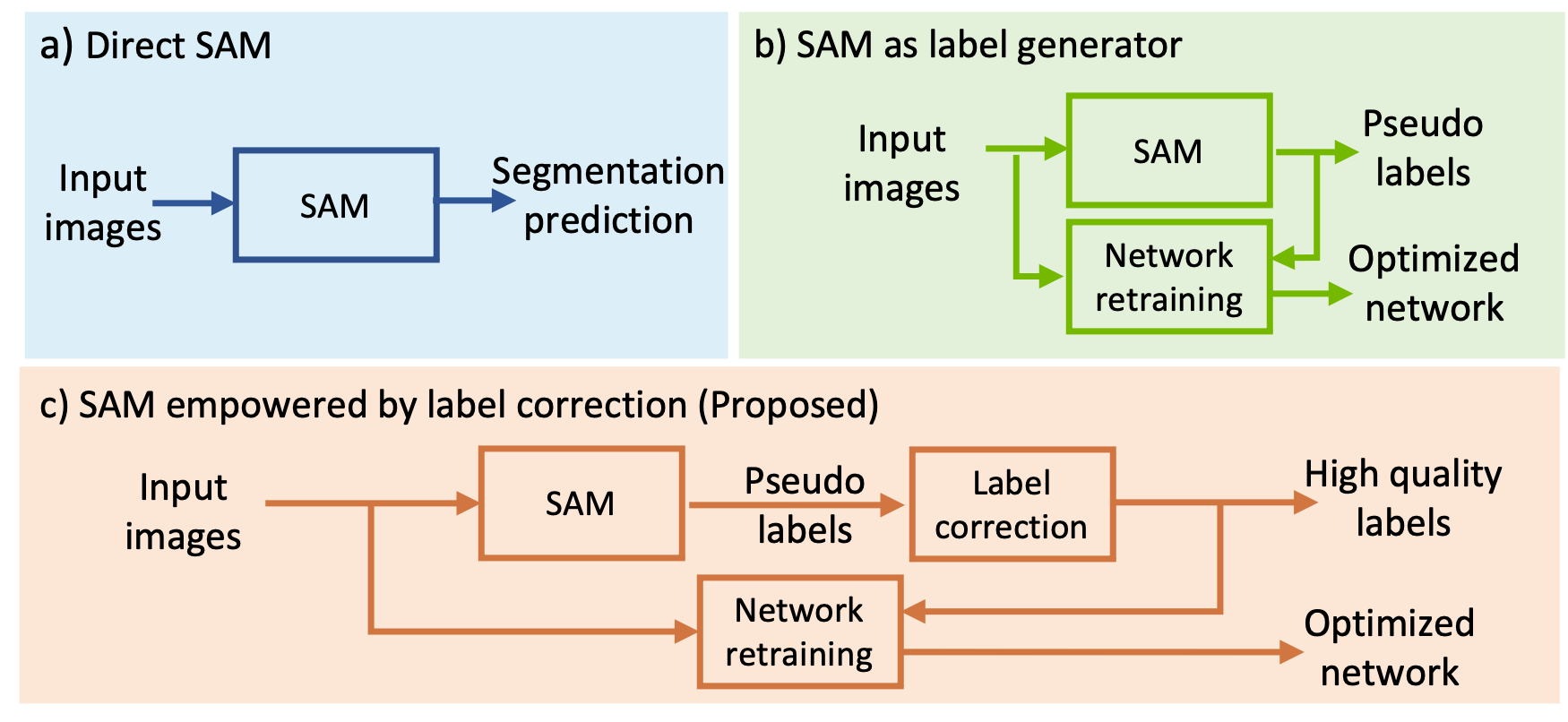}
\end{center}
\caption{A high-level comparison between existing SAM-based methods (a, b) and our proposed method (c).} %The framework includes a robust learning module and a self-correction module. Both modules work iteratively to correct noisy labels in training dataset and retrain the segmentation network.}

\label{fig:high_level}
\end{figure}

However, directly deploying SAM on medical segmentation tasks, the methods shown in Fig. \ref{fig:high_level}(a), has shown generally lower performance in comparison with state-of-the-art segmentation models like U-Net \citep{shi2023generalist, cheng2023sam, deng2023segment, he2023accuracy,roy2023sam}, despite its superior performance on natural images. Medical image segmentation presents unique challenges due to factors such as comparatively low signal-to-noise ratio, blurry boundaries, and salient features, making it a more intricate task compared to segmentation in natural image data. Consequently, pushing the limits of SAM in medical segmentation remains a particularly challenging endeavor.

In most of the medical datasets, there exists a significant performance gap between SAM segmentation and fully-supervised segmentation prediction. To bridge this gap, a new trend has emerged, wherein SAM is leveraged as a pseudo label generator for fully-supervised segmentation models \citep{zhang2023input, sun2023alternative, hu2023sam, li2023segment} or network finetuning \citep{cui2023all}, as shown in Fig. \ref{fig:high_level}(b). These approaches involve passing the data through a conventional segmentation network to hold on to the zero-shot nature of SAM, thus eliminating the need for pixel-wise annotations. Although the combination of SAM and fully supervised segmentation exhibits great promise, the performance is constrained by the quality of pseudo-labels generated by SAM, as the imperfect low-quality pseudo-labels can negatively impact the performance of the subsequent segmentation model, leading to lower segmentation performance.  % to combine SAM with supervised learning by using SAM to generate pseudo labels for the training of traditional segmentation model. 
%Due to the difference between natural data and medical data, the ability of SAM to segment medical images that demands professional knowledge is limited. Directly applying SAM to real-world medical scenarios generates unsatisfying results compared to state-of-the-art segmentation performance .
%Leveraging this foundational model to facilitate medical image segmentation still holds great promise. The masks generated by SAM can effectively provide coarse labels, serving as invaluable annotation tools for developing a precise and effective medical image segmentation model \citep{zhang2023input,hu2023sam,roy2023sam}. 

To address the above issues, we propose a label correction framework to push the boundary of SAM-based segmentation as shown in Fig. \ref{fig:high_level}(c). Inspired by the principle of self-supervised learning technique, our proposed framework can effectively correct pseudo-labels generated from SAM, improving the subsequent segmentation network to be resilient to imperfect annotation during the supervised training process and thus holding promise to improve their performance. A comparison between our proposed method and the existing SAM-based methods is shown in Fig. \ref{fig:high_level}. 

Within the domain of self-supervised learning against label noise (also called robust learning), existing advances mainly focused on image-level classification \citep{goldberger2016training,NEURIPS2018_ad554d8c,patrini2017making,jindal2016learning,sukhbaatar2015training,wang2020training,quan2020effective, zhang2020distilling}, which cannot be easily extended to segmentation tasks with pixel-wise predictions. On the pixel level, %our group developed a co-teaching framework to correct imperfect labels in lung segmentation\citep{huang2021coseg}. 
Wang et al.\citep{wang2020meta} assigned small weights on noisy pixels based on training an additional meta mask network. Zhang et al.\citep{zhang2020characterizing} applied a confident learning technique to characterize label noises from the training data to generate pixel-level noise identification maps. Zhu et al. \citep{zhu2019pick} proposed a framework
to examine the label quality and assign different weights to the noisy labels at the image level. The work \citep{mirikharaji2019learning} considered the spatial variations in the quality of pixel-level annotations to learn spatially adaptive weight maps and adjusted the
contribution of each pixel in the optimization of deep networks. Our previous work \citep{huang2021coseg} generalized the structure of co-teaching \citep{han2018co} into segmentation tasks by training two segmentation networks simultaneously to pick clean image samples for each one. However, a noteworthy limitation of those methods is that they rely on a subset of expensive expert-level clean labels during training. In practice, this is not feasible when working with SAM-generated pseudo-labels, as no expert-level supervision is involved in the process.

\begin{figure*}[htb]
\begin{center}
\includegraphics[width=0.9\textwidth]{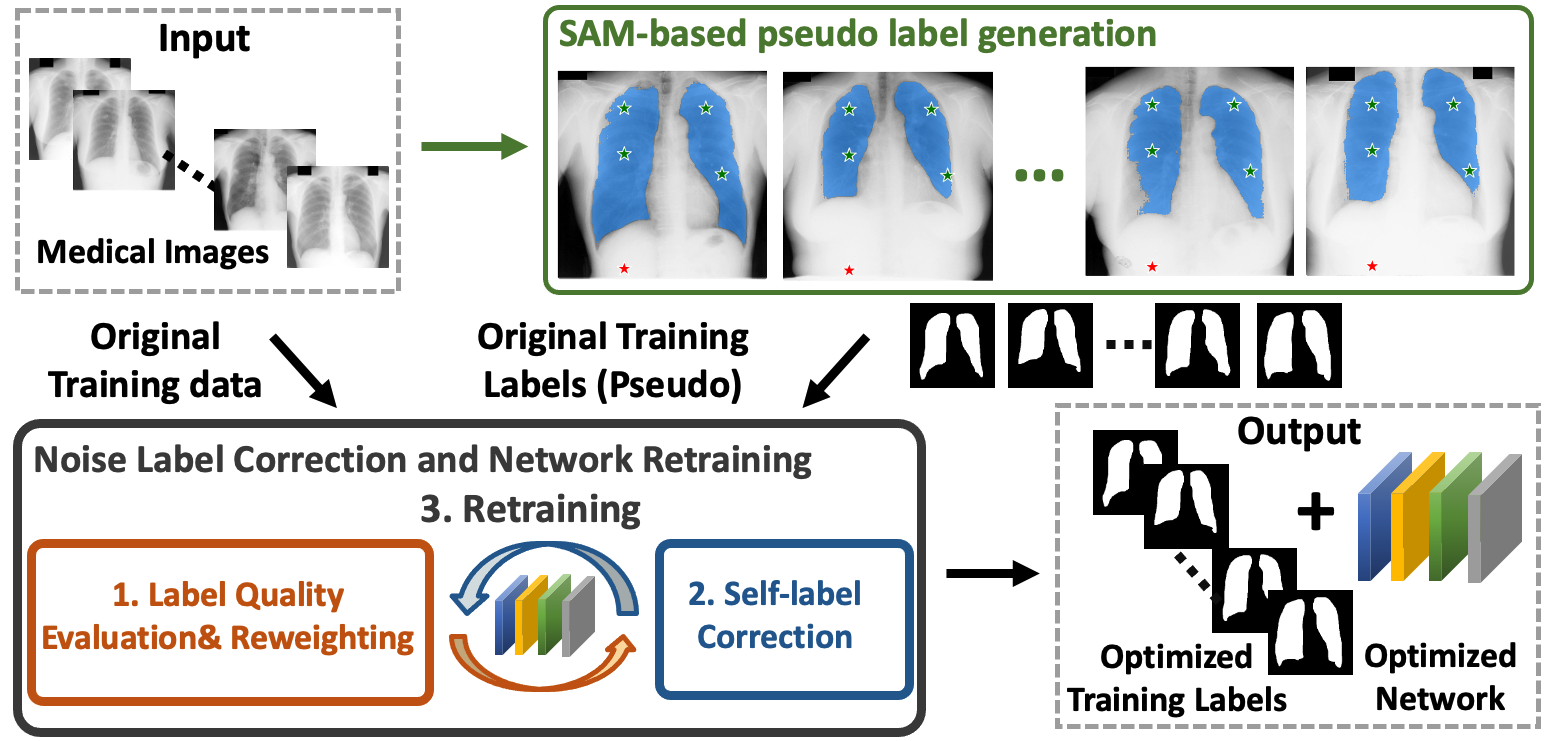}
\end{center}
\caption{Overview of proposed model. Given input medical images, our framework generates pseudo labels using SAM. The quality of pseudo labels is optimized for supervised learning. In SAM-based pseudo label generation, the green stars and red stars indicate positive and negative point prompts.} %The framework includes a robust learning module and a self-correction module. Both modules work iteratively to correct noisy labels in training dataset and retrain the segmentation network.}

\label{fig:top_flow}
\end{figure*}

In this paper, we propose a novel label correction framework for SAM-generated pseudo-label improvement. The framework consists of the following three modules: label quality evaluaiton and reweighting module, self-correction module, and retraining module. In the first module, we design a multi-level reweighting strategy to automatically justify the label quality at the image level, thus eliminating the need for predefined clean image-level labels in a separate set. Benefiting from it, our proposed approach does not require any subset of human-annotated clean labels or predefined high-fidelity labels. After getting the confidence prediction from the label quality evaluation module, the self-correction module automatically corrects the noisy labels, enriching the high-quality pseudo labels for further training. 
% \textcolor{red}{(YG:we do have clean label on pixel-wise, need to think about and revise. Emphasize on pre-defined clean label on pixel wise. No clean image involved)}
Finally, we design a retraining module so that the network can benefit from the whole training set. 

To the best of our knowledge, we propose the first label correction framework for SAM segmentation. Specifically, our main contributions include: 

(1) We propose a pseudo-label correction framework to push the boundary of SAM-based segmentation for medical images. The whole framework largely increases the performance of the SAM-based segmentation framework while holding the nature of zero-shot, i.e., no pixel-wised human-made annotation involved in training.

(2) We develop a multi-level reweighting strategy for robust training against low-quality labels without any assumptions required for human-annotated clean or predefined high-fidelity labels, which is a particular fit for SAM-generated pseudo labels. 

% \textcolor{red}{
% (3)We integrated our label correction framework with a high-performance weakly supervised learning framework which does not require any pixel-wise annotation in training. 
% }

% task \textcolor{red}{ without needing any clean samples.}. It can efficiently enhance the quality of the training set and the segmentation capability of the network. % presuming the existence of any clean samples
%(2) We perform a study on distinguishing ``hard"  from noisy labeled pixels for robust learning with overfitting control. % We rely on an uncertainty-based scheme to quantify and automatically correct for corrupted labels over iterations. %A novel pixel-wise reweighting policy is designed for robust learning and overfitting control without any prior knowledge of noise. In addition, we devise an uncertainty-based scheme to quantify and automatically correct corrupted labels over iterations. %% avoid the degeneration caused by memorizing noisy pixels and ignoring hard pixels.
(3) We experimentally demonstrate that our model significantly outperforms the existing SAM-based methods in three public datasets, boosting the performance of zero-shot learning towards fully supervised learning. %To the best of our 

%% file: sections/relatedwork.tex
\section{Related work}

% \textcolor{red}{YG: Need a thorough related work section to discuss 1) label correction in classification; 2) lable correction in segmentation
% 3) other weakly supervised learning framework in medical iamge segmentation}

\subsection{Segment Anything Model in Medical Segmentation}
 Due to the difference between natural data and medical data, the ability of SAM to segment medical images that demand professional knowledge is limited. Directly applying SAM to real-world medical scenarios generates unsatisfying results compared to state-of-the-art segmentation performance \citep{he2023accuracy, roy2023sam,ji2023segment,zhang2023segment,Chen2023SegmentAM, zhang2023input,cheng2023sam,deng2023segment}.
Leveraging this foundational model to facilitate medical image segmentation still holds great promise. The masks generated by SAM can effectively provide coarse labels, serving as invaluable annotation tools for developing a precise and effective medical image segmentation model \citep{zhang2023input, hu2023sam, li2023segment}. However, the performance of the segmentation is not ideal or close to fully-supervised learning.
% V{\'a}zquez et al.\citep{vazquez2021two} used a self-learning label correction module to estimate the chance of correctly labeled samples by comparing the similarity and density measures to each class. Co-correcting proposed by Liu et al. \citep{liu2021co} employed a dual network structure that only updated parameters when there is a mutual agreement on predictions, and utilized a label probabilistic modeling and curriculum-based label correction based on the distance map of features for each category. Wu et al. \citep{wu2022labcor} proposed a binary relevance-based label correction method that identifies the reliabilities of sample-wise base and meta-predictions by training a base-classifier and meta-classifier, and corrects the wrongly classified samples. Kremer et al. \citep{kremer2018robust} designed an active label correction method that featured noise model-dependent loss functions and maximum expected model change criterion to regularized risk functions. Yi et al. \citep{yi2019probabilistic} designed a noisy label handling structure to update and correct the label via back-propagation, which does not require prior knowledge of noise.
% Gong et al. \citep{gong2022synergistic} trained two networks with a joint loss of cross entropy loss and agreement loss to reduce the divergence between different classifiers, and then corrected the confident noisy samples iteratively.

\subsection{Label Correction in Segmentation}
Zhu et al. \citep{zhu2020fcn} proposed an organ segmentation model that used atlases to estimate the label deformation and used a fully convolution network to learn the relationship between image patches and measure the confidence of each patch and correct wrong labels.
A pseudo-label correction framework that corrects the label at pixel-level and image-level was employed by Wu et al. \citep{wu2022pseudo}, and this framework can correct the falsely predicted pixel labels and missed predicted pixel labels by a two-branch architecture, so as to improve the recall and precision.
Yi et al. \citep{yi2021learning} addressed pixel-level noisy label segmentation problem by graph-based label noise detection and correction framework, where the spatial adjacency and semantic similarity constraints are used to construct superpixel-based graph of the image and the clean labels are propagated to the noisy labels using a graph attention network. Ibrahim et al. \citep{ibrahim2020semi} used the ancillary model and primary model with weak annotation and strong annotation, and fused the features from two models by self-correcting module. 
A similar idea of early learning and label correction was further explored by Feng et al. \citep{feng2023weakly}, where a unified inflection hyperparameter was used to decide the label updating timing along with the training of an exponential moving average model. However, those methods are not optimized for medical image analysis.

\subsection{Self-supervised Learning in Medical Image Segmentation}

Supervised segmentation deep learning poses the need for labeled training data, which is costly in the medical image segmentation field, as domain expertise and large data scales are highly preferred. 
Self-supervised learning can accommodate this issue by training on unlabeled data.
Kalapos et al. \citep{kalapos2023self} investigated the transfer learning capability of self-supervised pretraining approaches for the downstream medical image segmentation tasks. Ouyang et al. \citep{ouyang2022self} proposed a superpixel-based self-learning strategy and an adaptive local prototype pooling network that performed well in the few-shot medical image segmentation. Recently, Liu et al. \citep{liu2022adaptive} exploited the early learning behavior of semantic segmentation under noisy pixel-level labels and proposed a label correction strategy that is adaptive to early learning to improve the segmentation performance.

%% file: sections/method.tex
\section{Methodology}

Our goal is to automatically correct imperfect pseudo labels generated from SAM in training a segmentation network. As illustrated in Fig. \ref{fig:top_flow}, pseudo labels are generated using SAM. Then, our proposed method contains three modules to detect and correct low-quality labels for network training: (1) label quality evaluation and reweighting, (2) self-correction, and (3) retraining. Step (1) trains an initial segmentation network by weighing an subset of high-quality pixel and labels. Step (2) enriches the size of reliable dataset by correcting low-quality pixel to high-quality, namely label correction. Step (3) retrains the segmentation network by the improved dataset with updated labels.

% Based on the confidence prediction, the self-correction module automatically corrects the noisy labels, enriching clean labels for further training. Considering the representation learning as an iterative process, we further design an iterative learning framework so that the two modules can be jointly improved over iterations.

%For instance, a hard pixel may be labeled as class A 47 times and class B 43 times if it is labeled 100 times independently by an expert. 
%This type of pixels usually appears around the boundary of multiple classes.

\begin{figure*}[htb]
\begin{center}
\includegraphics[width=1\textwidth]{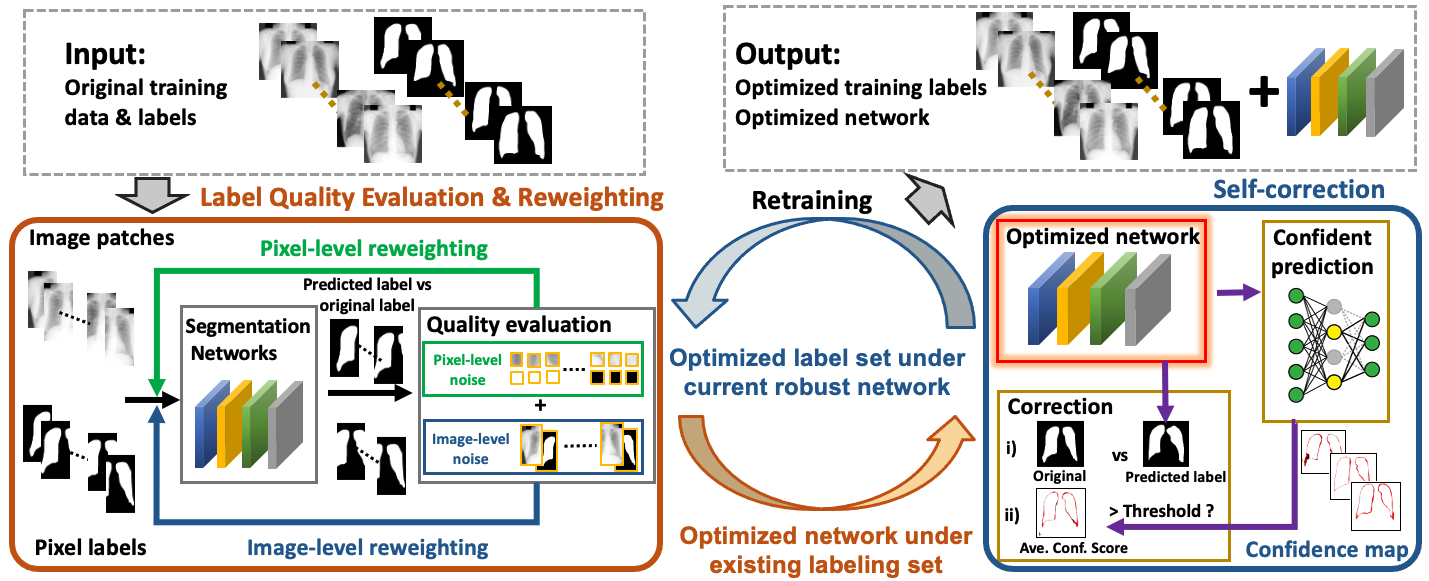}
\end{center}

\caption{The system flow of our proposed label correction model. Given training data and pseudo labels, our \textit{label quality evaluation and reweighting} module adopts a multi-level reweighting strategy at both image-level and pixel level to robustly train the network against noisy labels. Based on the confidence prediction, the \textit{self-correction} module corrects low-quality labels to enrich the clean set for further training. Benefiting from those two modules, a new network is retrained using the updated labels.} %The framework includes a robust learning module and a self-correction module. Both modules work iteratively to correct noisy labels in training dataset and retrain the segmentation network.}

\label{flow}
\end{figure*}

\subsection{Segment Anything Model in Label Generation}\label{sammethod}
We choose to utilize the existing SAM model \citep{kirillov2023segment} to generate pseudo labels for the training of a supervised segmentation network. The SAM model comprises three main components. Firstly, for the image encoder, Vision Transformer \citep{dosovitskiy2020image} pretrained with mean absolute error (MAE) is used to process the input image. Secondly, a prompt encoder is employed to process sparse and dense prompts. The stars, as shown in Fig. \ref{fig:top_flow}, are represented by positional encodings, and the masks are embedded by convolutions. Lastly, the mask decoder consists of a modified transformer decoder block and a dynamic prediction head. 

To facilitate effective relation exploration between prompts and prompt-image, SAM employs prompt self-attention and prompt-image two-way cross-attention mechanisms, which subsequently update the embeddings. The output token is then mapped to the dynamic linear classifier by maximum a posterior (MAP) estimation after two decoder blocks, resulting in the output of the mask foreground probability. Consequently, a pseudo label is generated considering the tissue region in medical images.

% The model is built on the largest segmentation dataset which has 11 million high-quality natural images and over 1 billion segmentation masks. The original release of SAM includes the evaluation of 23 datasets of natural images.
%By gradually pulling away data with low-quality label,
% the rest of data will be used to train a comparatively robust model.

\subsection{Label Quality Evaluation and Reweighting Strategy}\label{RL}
\label{sec:robust training}

\textbf{Definition of Noisy Labels.} In medical segmentation, we term image with labels that have high fidelity with expensive and expert-level annotations (i.e., ground truth) as \textit{clean image samples} and those pixels with labels that agree with expert-level annotation as \textit{clean pixels}. Meanwhile, we term image samples that contain a subset of cheap unreliable annotations as \textit{noisy image samples} and name those mis-labeled pixels (i.e., with labels different from ground truth) as \textit{noisy pixels}. Note that in SAM-generated pseudo labels, there is no clean image label as SAM can not achieve a perfect segmentation performance with 100\% accuracy \citep{cheng2023sam, deng2023segment, he2023accuracy,roy2023sam}. The number of noisy pixels within an image defines its ``purity'' at the image level.

In the following proposed framework, we establish a multi-level strategy to evaluate and reweight noisy labels simultaneously at image and pixel levels. Such strategy was designed to address noisy label caused by the nature of SAM image generation and consistent morphological similarities present across medical images in segmentation tasks. In comparison with existing methods, we add image-level reweighting for two reasons. First,  SAM generates pseudo labels on image-level basis. There are image frames that are generally underperformed in SAM-based label generation (example referred to Fig. \ref{fig:3D}(c)). These frames may lead to the presence of low-quality noisy labels surrounding individual clean pixels. The learning of local features and labels could mislead the network training. Secondly, in the context of medical segmentation tasks, there exists a morphological structural resemblance among various medical images. For instance, in lung X-ray images, both lung lobes are consistently positioned in the center. The noisy level at image level indicates how the network performs at similar segmentation regions at a macro-scale. Thus, it is important to have image level constraint on the cross image similarity in addition to pixel level-based reweighting.

\textbf{Label Quality Evaluation.} We evaluate the quality of training labels and detect low-quality label annotation using a segmentation neural network, as shown in Fig. \ref{flow}. During the training phase, avoiding training on noisy pixels and noisy image samples can help to improve the robustness of the network against label corruption. Thus, in this module, we apply both image-level and pixel-level schemes for label quality evaluation to avoid performance degeneration. Considering that labels from clean pixels should be similar to the ground truth in terms of semantic structure, but noisy pixels do not 
%\textcolor{red}{(YG:emphasize on semantic  structure that is similar)}
, our module leverages cross entropy (CE) loss for noisy label detection and evaluation since it assesses the probabilistic similarity between the actual label and the predicted label. More specifically, pixels with large CE losses are likely to be noisy pixels, and image samples with large CE losses are likely to be low-quality. Therefore, we choose to detect and lower the weights of such pixels and images in the training procedure to help networks focus on learning from clean pixels and purer image samples. In particular, if we denote $\Omega$ as an image sample, the CE losses for an image sample $\Omega$ and a pixel $x$ where $x\in \Omega$ are defined as follows, respectively: 
% \begin{equation}\label{CE}
%     L_{CE}(x)=- \sum_{l=1}^{k} g_l(x)\log(p_l(x)), \quad L_{CE}(\Omega)=\sum_{x\in\Omega} L_{CE}(x)
% \end{equation}
% %\begin{equation}\label{CE2}
% %    L_{CE}(\Omega)=\sum_{x\in\Omega} L_{CE}(x)
% %\end{equation}

\begin{equation}\label{CE2}
   L_{CE}(\Omega)=\sum_{x\in\Omega} L_{CE}(x)
\end{equation}
\begin{equation}\label{CE}
    L_{CE}(x)=- \sum_{l=1}^{k} g_l(x)\log(p_l(x))
\end{equation}
where $k$ is the number of total classes, $p_l(x)$ provides the predicted probability of pixel $x$ belonging to class $l$, and $g_l(x)$ is the one hot label for $x$.

\textbf{Reweighting Strategy.} Upon the detection of low-quality labels, we leverage a multi-level reweighting strategy to assign image samples and pixels with different weights. This multi-level reweighting strategy is designed to provide a multi-scale analysis on SAM-generated medical image masks.
%Pixel-level noisy label and image-level noisy label both exist in medical image segmentation tasks. 
In particular, our reweighting strategy measures the pixel-level and image-level noises simultaneously.
% \textcolor{red}{(YG:why to choose multi-level weight. Summerize the difference between ours and PINT paper, discuss in our next meeting)}. 
Along with the reweighting strategy, our total loss function is a combination of the pixel-wise weighted CE loss $L_{CE}$, the dice loss $L_{Dice}$, and the $L_2$-regularization term on the parameters $W$ of the network: 
% \begin{equation} \label{loss}
\begin{multline} \label{loss}
    L_{total} = \sum_{\Omega} \lambda(\Omega) \Big(\sum_{x\in \Omega} \alpha(x) L_{CE}(x) + L_{Dice}(\Omega)\Big) \\
    + ||W||_2^2,
\end{multline}
% \end{equation}
\begin{equation}
L_{Dice}(\Omega) = 1- \frac{1}{k} \sum_{l=1}^{k}\frac{2\sum\limits_{x \in \Omega}(p_l(x)g_l(x))}{\sum\limits_{x \in \Omega}(p_l(x))^2  + \sum\limits_{x \in \Omega}(g_l(x))^2},
\end{equation}
where $\alpha(x)$ is the pixel-wise weight assigned to the pixel $x$ and $\lambda(\Omega)$ is the image-wise weight assigned to the image $\Omega$. These weights are obtained from our novel multi-level reweighting strategy as described below. %It helps to distinguish the noisy pixels from the hard pixels.First we initialize all weight values equal to 1, i.e., $\alpha(x)=1, \ \forall x$ and $\lambda_1(\Omega)=1, \ \forall \Omega$. 

Starting from the $E_{start}$-th epoch, we begin to assign pixel-wise weights $\alpha(x)$ and image-wise weights $\lambda(\Omega)$ on the training set to avoid overfitting to noisy labels. Intuitively, for an image sample that is severely low-quality with noisy pixels, the best way to avoid performance degeneration is to discard it. Thus, in the image-level reweighting scheme, we ignore the entire image sample $\Omega$ if it has a very large CE loss. The proportion of samples ignored in the training set is controlled by the forget rate, $\beta$. That is, in each mini-batch, we only select $1-\beta$ percentage of samples with the smallest CE losses for network optimization. 

%$L_{CE}(\Omega)>q_0$ is larger than a certain threshold, then we  i.e., set $\lambda_1(\Omega)=0$ in (\ref{loss}). Here we set $q_0$ to be the $1-\beta_0$ quantile of the set $\{L_{CE}(\Omega): \Omega \text{ is a sample in the training set}\}$. In other words, we omit the $\beta_0$ proportion of samples with the largest CE losses.  
For datasets containing a large number of noisy image samples, simply ignoring all of them will lead to severe overfitting, especially in biomedical applications. Thus, we further introduce a novel pixel-level reweighting scheme to assign different weights to each pixel. For a given noisy image sample $\Omega$, pixels with large CE losses are more likely to be noisy pixels. Thus, we set $\alpha(x)=0$ in Equation (\ref{loss}) if $L_{CE}(x)>q$, where $q$ is the $1-\gamma$ quantile of the set $\{L_{CE}(x): x\in\Omega\}$, 
% $$(\textcolor{red}{ a new parameter $\gamma$ here is the quantile parameter, and it is $1-\gamma=0.95$ in the code. $\beta$ was used here before and is duplicated with the forget rate.})
and $\alpha(x)=1$ if $L_{CE}(x)\leq q$. The image-wise weight $\lambda(\Omega)=1$ if the $L_{CE}(\Omega)$ ranks in the $1-\beta$ smallest CE losses, otherwise  $\lambda(\Omega)=0$.
Then, our network is optimized by the reweighted loss function, and on the next epoch, we recalculate the weights based on the above strategy. %ote that different from Co-Seg,  we add an additional pixel-level reweighting scheme. Benefiting from our two-level reweighting strategy, we only need to train a single network (instead of two networks in \citep{huang2021coseg}) to reduce the training load (time, parameters and so on).  

\begin{figure*}[t]
\begin{center}
\includegraphics[width=0.85\textwidth]{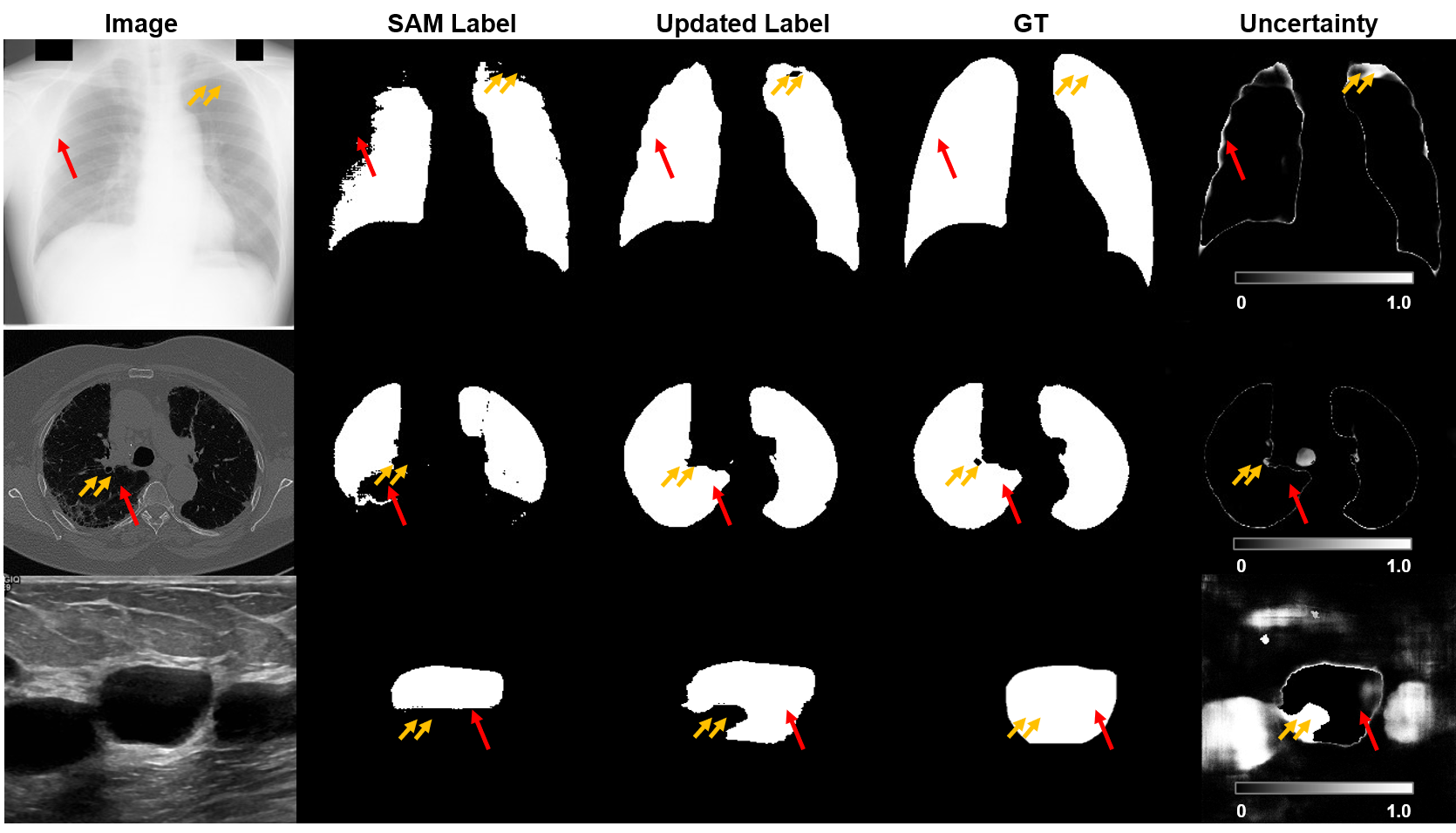}
\end{center}
\caption{Representative example of pseudo labels correction from both JSRT dataset (first row), lung CT dataset (second row), and BUSI dataset (third row). In each row, updated labels using our proposed method significantly convert noisy pixel labels to clean pixel labels. The red arrow highlights the corrected region. The yellow double arrows indicate a region with high uncertainty, leading to missed label updates.} 
\label{fig:sam_label_updated}
\end{figure*}

\subsection{Self-Correction Module}\label{CLC}
\textbf{Confidence Prediction.}
% \textcolor{red}{(YG:expand with more details and equations)} 
In the self-correction module, we selectively correct a subset of noisy labels based on confidence measurement in \citep{gal2016}. 
Inspired by \citep{hu2019supervised,sedai2018joint}, we use the dropout-based Monte Carlo sampling approach to access the pixel-wise confidence information. 
In particular, we turn on dropout before each convolution layer and repeat $Num$ stochastic forward passes of each training image through the network. In the $j$-th ($0<j\leq Num$) forward pass, we compute the prediction vector $v^{x} = (v^{x}_{1}, v^{x}_{2}, ..., v^{x}_{Num})$ for each input pixel $x$.
If the prediction of the $x$ pixel coincides with the prediction of the $x$ pixel using the network without dropout, then $v^{x}_{j} = 1$, otherwise $v^{x}_{j} = 0$. The confidence score $CS^x$ is then defined as 
\begin{equation}
    CS^{x} = \frac{\sum_{j = 1}^{Num}v_{j}^{x}}{Num}
\end{equation}
Noticing that $CS$ is based on element-wise computing on a pixel $x$, a confidence map can be formed for each training image, as shown in Fig. \ref{flow}.

\textbf{Self-Correction.} The self-correction module aims to enrich the training dataset with more high-quality pseudo labels. A label in the training set is corrected only when the following two conditions are satisfied simultaneously: 1) there is a mismatch between the predicted and original label, and 2) the predicted label has a high confidence score. To avoid the case where pixels are mis-corrected, we rely upon the confidence score, $CS^x$, to justify whether we need to avoid the potential mis-correct of pixels. Only pixels that we confidently know the correct labels will be updated in the new training dataset. We expect the pixels with updated labels will be assigned a higher weight in future reweighting and retraining, thus providing better guidance in supervised learning. 

% The presence of intricate image structures, like the delicate boundaries of organs, gives rise to "hard" pixels that raise a significant challenge for the networks to predict accurately. The ground truth annotations of these complex regions often present limited fidelity.
% In the self-correction module, with the conditions met, we will not mis-correct the ''hard" pixels. Those pixels still remain in the training set for future discriminative learning in the retraining module. 

% \textcolor{red}{(YG:why you do not correct hard pixel? Possible answer: hard pixel is from due to structure of the image. The ground truth itself is with low-fiedelity)}

\subsection{Retraining Module}  

After implementing label corrections, we re-input the updated training labels to a randomly initialized segmentation network and start a new training procedure. We will repeat the label quality evaluation and reweighting from the updated training dataset. A new parameter network, $\hat{W}$, will be output after the retraining. Similarly, this retraining procedure uses the loss function defined in Equation (\ref{loss}) by considering both image and pixel level. Then, this segmentation network, $\hat{W}$, will be used to make the final prediction for any testing data.%the final segmentation prediction is obtained from this network.  %consisting of robust learning module (Section \ref{RL}) and self-correction module (Section \ref{CLC}). This procedure repeats until segmentation performances on the training dataset reaches convergence.

%Based on the updated dataset, we retrain a final segmentation network, which shares the same network structure as one of the peer networks. Similarly, the retraining uses loss function defined in Eq. \ref{all_loss}. Then, this network will be used for final predictions.%the final segmentation prediction is obtained from this network. 

%% file: sections/results.tex
\section{Experiments}

\subsection{Datasets}

We conduct evaluations on three publicly available datasets: the X-ray dataset from the Japanese Society of Radiological Technology (JSRT) \citep{shiraishi2000development}, lung CT from the Open Source Imaging Consortium (OSIC) Kaggle dataset \citep{lungCT2020}, and the Breast Ultrasound Images (BUSI) dataset \citep{al2020dataset}.

The JSRT dataset consists of 247 grayscale chest radiographs. The ground-truth lung masks are obtained from the Serial Chest Radiographs (SCR) database \citep{van2006segmentation}. Each chest radiograph is a 2D image with a dimension of $1024 \times 1024$ pixels. Following the work in \citep{he2019non,hwang2017accurate}, our training set and testing sets are split by the ID number. Thus, our training set contains 124 samples with odd numbers and the testing set contains 123 even numbered chest radiographs.

% 1.(The lung CT dataset consists of 111 volumes of CT scans. 
% Fifty volumes were chosen where, in at least one axial slice, the lung labels exceed more than 20\% of the pixels.)
% Or we can say: 2. (

The lung CT dataset used in this work comprises 50 volumes of CT scans that primarily focus on the lung region.
%)\textcolor{red}{New information added above. (because not all the 111 volumes contain the lung regions in decent sizes. The other 61 volumes have lung regions only in a small portion of every slice. keeping these images would make SAM hard to generate pseudo labels, one of the reasons is that the point prompts are not valid.)} 
The ground truth is provided from the OSIC pulmonary fibrosis dataset \citep{li2022research}. Each CT scan is a 3D volume with a dimension of 768 $\times$ 768 pixels or 512 $\times$ 512 pixels in the axial plane, and the number of axial slices ranges from 17 to 408 among scans. In this work, each CT scan is treated as an independent sample. We validate the performance using a cross-validation setup by randomly splitting the whole dataset by volumes into two groups and rotating the training and testing between the two groups. Averaged results in cross-validation are reported. 

The BUSI dataset \citep{WalidAlDhabyani.2020} includes 780 ultrasound images depicting normal, benign, and malignant cases, each with corresponding segmentation maps. The images are acquired from 600 female patients, with an average size of 500 $\times$ 500 pixels. For our study, we exclude images with landmarks and texts and utilize a subset of this dataset that consists of 647 benign and mailgnant images depicting cases of breast tumors. We randomly select 520 images for training and 127 images for testing.

\subsection{Implementation Details.}
In JSRT data, we resize all images from $1024 \times 1024$ pixels into $256 \times 256$ pixels and crop each image into small image patches with $256 \times 128$ pixels for data augmentation. 
In lung CT data, we resize the images from $768 \times 768$ or $512 \times 512$ into $256 \times 256$ pixels. Similarly, we crop each image into small image patches with $256 \times 128$ pixels. In BUSI dataset, we resize benign and malignant images to $256 \times 256$ pixels. %Because we have enough number of images, it is not necessary to patchify the data into small pieces. 
During the training, we used Adam optimizer\citep{kingma2014adam}. The learning rate was initialized as 0.001 with the $\beta_1$ set as 0.9, and the $\beta_2$ set as 0.999. The learning rate was scheduled by a dynamic decay of the percentage of the rest epochs after 10 epochs. The $\beta_1$ was set as 0.1 after 10 epochs. In total, the networks were trained 300 epochs to ensure convergence. The quantile parameter was set as 0.05. We empirically set forget rate $\beta$ as 20\% for JSRT, 10\% for lung CT data,  and 10\% for BUSI. We set the dropout number $N$ as 100 for JSRT, 120 for lung CT data, and 100 for BUSI. The experiments were carried out in parallel on two RTX A6000 GPUs.

\subsection{Label Correction Performance}
We first evaluate the performance of label correction. In Fig. \ref{fig:sam_label_updated}, we present three representative examples, from the JSRT dataset (first row), the lung CT dataset (second row), and BUSI dataset (third row) respectively, to show the label correction performance of our proposed model from the SAM-generated pseudo labels (denoted as SAM Label). 
Compared with the original pseudo labels, the corrected labels are more similar to the ground truth, showing a significant improvement in label quality. In addition, the boundaries get smoother, and the regions with missing pixels, where the red arrow points in each figure, are automatically corrected. 
% As can be seen in Fig. \ref{fig:sam_label_updated}, the confidence maps show higher values at the location of noisy pixels, which are consistent with the noisy pixels marked in the noisy samples. 
In concordance with the results in \citep{kendall2015bayesian}, the confidence maps (denoted as uncertainty) in Fig. \ref{fig:sam_label_updated} have higher uncertainty among the tissue boundaries, as shown in double arrows. 
It evidently shows that the confidence maps could accurately locate the hard pixels caused by the boundary ambiguity. Thus, our label correction module will not miscorrect those hard pixels which are difficult for the networks to make predictions. 

Table \ref{tab:sam_labelcorrection} quantitatively compares the improvement of label quality. Against the ground truth, we compare the percentage of pixels that are clean in SAM-generated labels and updated labels. True positive (TP), false positive (FP), true negative (TN), and false negative (FN) are all calculated and compared. True positive rate is increased by $3.79\%$ in JSRT, $4.55\%$ in lung CT, and $3.54\%$ in BUSI. This is considered a significant improvement because most of the updated labels occur at the boundary of the lung region, as shown in Fig \ref{fig:sam_label_updated}. Moreover, we observe the decrease of FP and FN with TN increase as well.   
Our superior performance indicates that our model can push the limit of label quality, providing high-quality labels for the training of the segmentation network. It is worth mentioning that we observe a higher FP rate in the BUSI dataset compared to the JSRT and CT dataset, which is consistent with \citep{Shen.2021} where a high false positive rate for the BUSI dataset is reported due to a comparative low signal-to-noise ratio (SNR).

\begin{table}[t!]
\footnotesize
\centering
\caption{Label correction performance for SAM noise labels for JSRT, CT, and Ultrasound datasets.}
\begin{tabular}{cccccc}
\hline
                      &               & TP      & FP     & TN      & FN     \\ \hline
\multirow{3}{*}{JSRT} & SAM           & 81.58\% & 0.58\% & 99.42\% & 18.41\% \\
                      & Updated label & 85.37\% & 0.14\% & 99.86\% & 14.62\% \\
                      & Difference    & 3.79\%$\uparrow$  & 0.44\%$\downarrow$ & 0.44\%$\uparrow$  & 3.79\%$\downarrow$ \\ \hline
\multirow{3}{*}{CT}   & SAM           & 88.57\% & 0.90\% & 99.10\% & 11.43\% \\
                      & Updated label & 93.12\% & 0.64\% & 99.36\% & 6.88\% \\
                      & Difference    & 4.55\%$\uparrow$  & 0.26\%$\downarrow$ & 0.26\%$\uparrow$  & 4.55\%$\downarrow$ \\ \hline

\multirow{3}{*}{BUSI}   & SAM           & 61.43\% & 4.35\% & 95.64\% & 38.57\% \\
                      & Updated label & 64.97\% & 4.07\% & 95.92\% & 35.03\% \\
                      & Difference    & 3.54\%$\uparrow$  & 0.28\%$\downarrow$ & 0.28\%$\uparrow$  & 3.54\%$\downarrow$ \\ \hline
\end{tabular}
\label{tab:sam_labelcorrection}
\end{table}

\begin{figure*}[ht!]
\begin{center}
\includegraphics[width=0.88\textwidth]{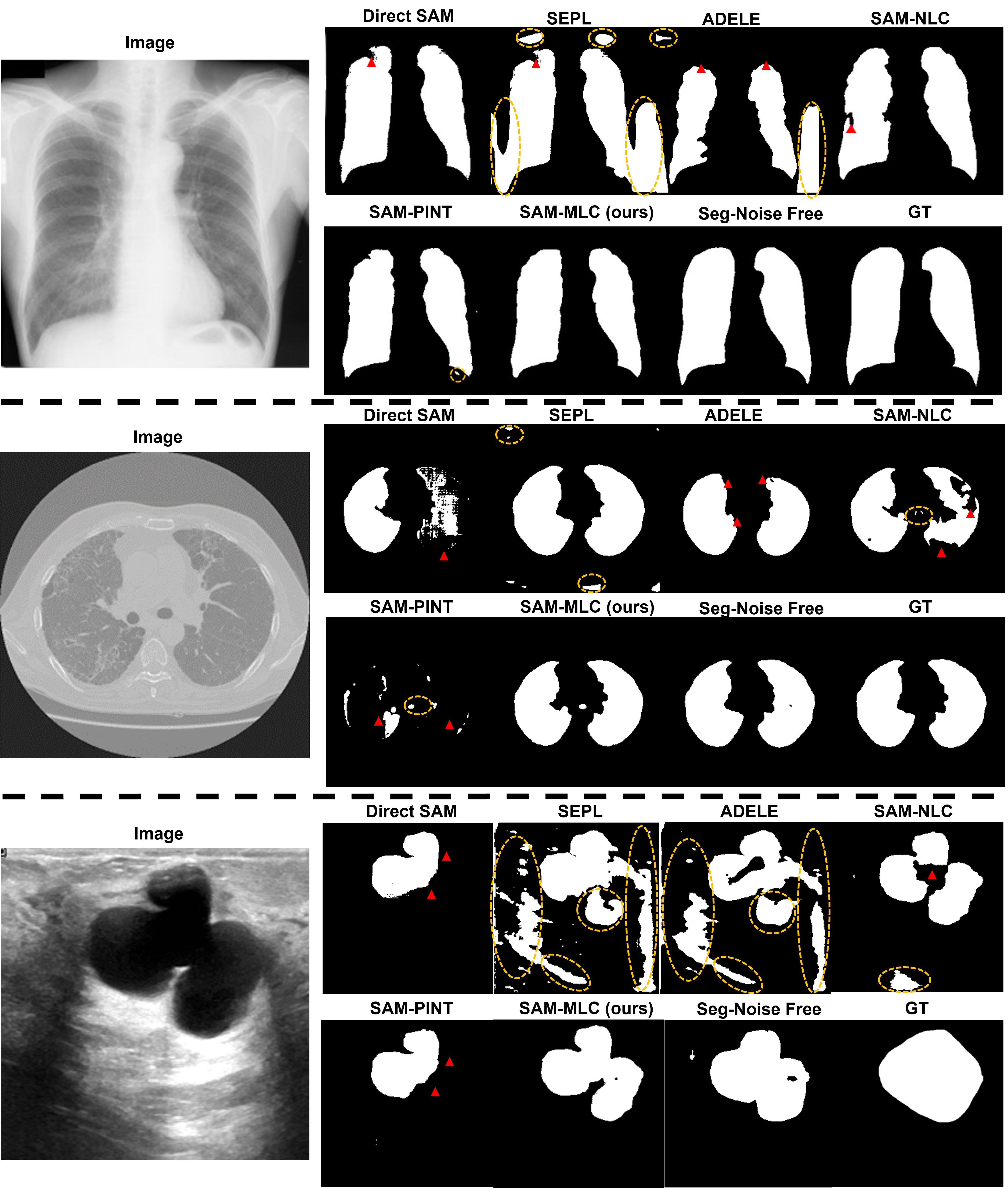}
\end{center}
\caption{Segmentation performance of the baseline methods and the proposed method from the JSRT dataset (first row), lung CT dataset (second row), and BUSI dataset (third row). With the multi-level label correction framework, our proposed method (SAM-MLC) outperforms the existing label correction method (SEPL, ADELE, and SAM-PINT) and achieves comparable segmentation performance to the fully supervised learning network from noise-free dataset (Seg-Noise Free). Red triangles indicate representative regions that are misclassified as background. Gold dash circles indicate representative regions that are misclassified as tissue regions.} 
\label{fig:retrain_sam_figure}
\end{figure*}

\begin{table*}[ht!]
\footnotesize
\centering
\caption{Evaluation metrics of different models on three dataset with noisy labels from SAM. Best results are in \textbf{bold}. The second best performance is \underline{underlined}.}
\label{tab:retrain_sam}
\begin{tblr}{
  cells = {c},
  cell{1}{1} = {c=2,r=2}{},
  cell{1}{3} = {c=2}{},
  cell{1}{5} = {c=2}{},
  cell{1}{7} = {c=2}{},
  cell{3}{1} = {r=2}{},
  cell{5}{1} = {r=2}{},
  cell{7}{1} = {r=2}{},
  cell{9}{1} = {r=2}{},
  cell{11}{1} = {r=2}{},
  cell{13}{1} = {r=2}{},
  cell{15}{1} = {r=2}{},
  hline{1,3,5,7,9,11,13,15,17} = {-}{},
}
                &        & JSRT           &                & Lung CT        &                & BUSI           &                \\
                &        & Acc            & Dice           & Acc            & Dice           & Acc            & Dice           \\
{Direct SAM \citep{kirillov2023segment}}      & Tissue & 80.95\%          & 88.50\%          & 88.26\%          & 91.44\%          & 61.70\%          & 64.78\%          \\
               & Background   & 90.12\%         & 95.53\%          & 93.45\%          & 97.58\%          & 78.77\%          & 95.17\%          \\
{SEPL   \citep{Chen2023SegmentAM}}         & Tissue & 63.49\%          & 69.83\%          & 85.79\%          & 88.82\%          & 62.78\%          & 22.17\%          \\
                & Background   & 81.58\%          & 92.74\%          & 92.49\%          & 97.55\%          & 69.89\%          & 84.08\%          \\
{ADELE \citep{liu2022adaptive}}           & Tissue & 59.81\%          & 64.33\%          & 88.80\%          & 90.04\%          & 60.42\%          & 24.44\%          \\
                & Background   & 74.61\%          & 86.32\%          & 92.49\%          & 96.65\%          & 70.44\%          & 86.25\%          \\
{SAM-NLC \citep{jiang2023segment}}         & Tissue & 78.71\%          & 87.41\%          & 87.30\%          & 91.02\%          & 46.87\%          & 51.02\%          \\
                & Background   & 89.18\%          & 95.36\%          & 93.19\%          & 97.71\%          & 72.95\%          & 95.69\%          \\
{SAM-PINT \citep{shi2021distilling}}        & Tissue & 81.68\%          & 89.07\%          & 88.01\%          & 92.29\%          & 48.12\%          & 52.08\%          \\
                & Background   & 90.62\%          & 95.96\%          & 93.61\%          & 98.01\%         & 72.95\%          & 95.66\%          \\
{SAM-MLC (ours)}  & Tissue & \uline{85.51\%}  & \uline{91.88\%}  & \uline{92.70\%}  & \uline{94.67\%}  & \uline{66.04\%}  & \uline{68.26\%}  \\
                & Background  & \uline{92.65\%}  & \uline{96.80\%}  & \uline{95.98\%}  & \uline{98.55\%}  & \uline{81.42\%}  & \uline{96.09\%}  \\
Seg-Noise Free  (upper limit) & Tissue & \textbf{97.96\%} & \textbf{97.26\%} & \textbf{95.49\%} & \textbf{95.93\%} & \textbf{76.24\%} & \textbf{75.41\%} \\
                & Background   & \textbf{98.61\%} & \textbf{98.92\%} & \textbf{97.28\%} & \textbf{98.85\%} & \textbf{87.25\%} & \textbf{97.19\%} 
\end{tblr}
\end{table*}

\begin{figure*}[h!]
\begin{center}
\includegraphics[width=0.9\textwidth]{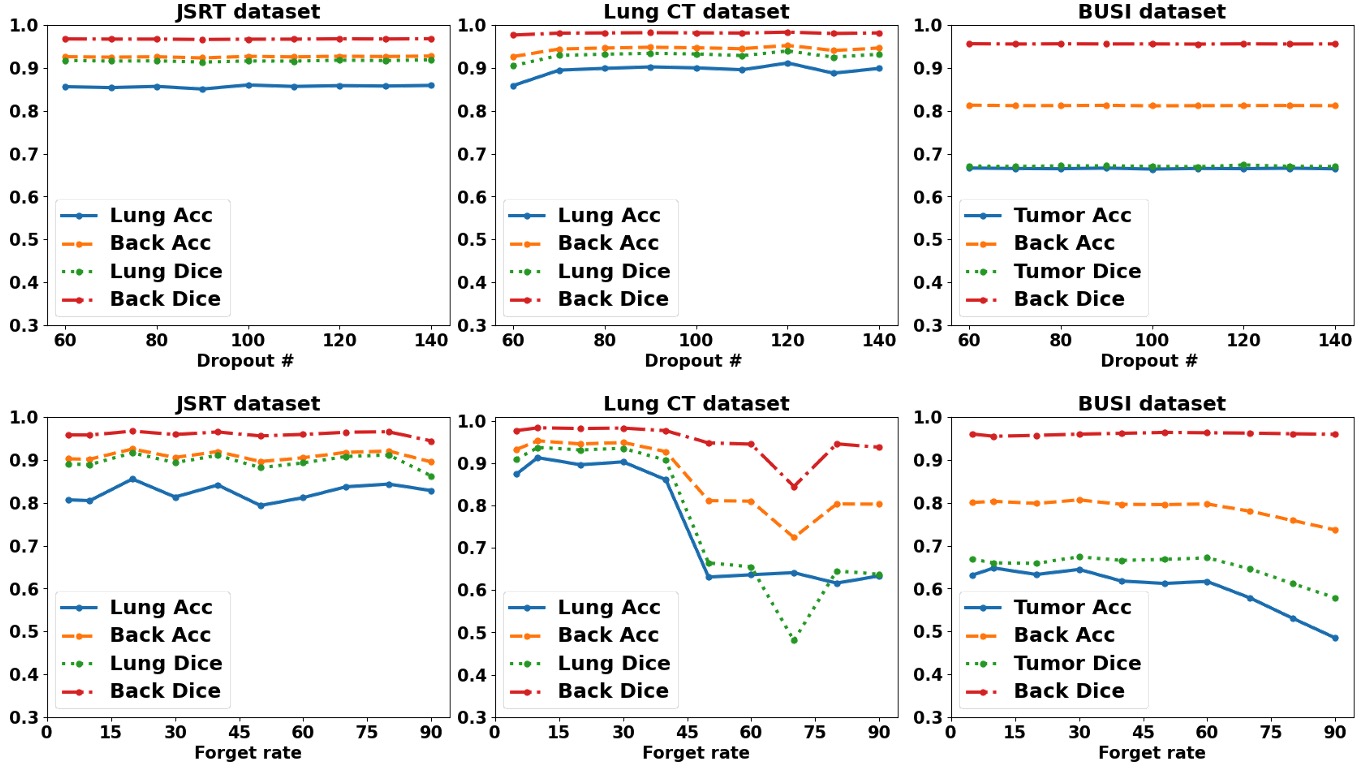}
\end{center}
\caption{Effect of hyperparameter on the segmentation performance in JSRT dataset, lung CT dataset, and BUSI dataset. Forget rate $\beta$ and dropout number $N$ are evaluated. Lung Acc = Lung Accuracy; Back Acc = Background Accuracy; Tumor Acc = Tumor Accuracy; Lung Dic = Lung Dice Coefficient; Back Dice = Background Dice Coefficient. } 
\label{parameter}
\end{figure*}

\subsection{Comparative Studies on Segmentation Performance}
We further evaluate the impact of label correction on prediction of segmentation task before and after retraining with updated labels. In particular, we compare the segmentation performance on the same testing dataset from the following methods: 1) Zero-shot segmentation directly from SAM (denoted as Direct SAM) \citep{kirillov2023segment}; 2) Segmentation using SAM Enhanced Pseudo Labels (SEPL) \citep{Chen2023SegmentAM}; 3) Segmentation using adaptive early-learning correction (ADELE) \citep{liu2022adaptive}; 4) Segmentation using pseudo labels without label correction (SAM-NLC) \citep{jiang2023segment}; 5) Segmentation using pseudo label corrected from existing robust learning method, pixel-wise and image-level noise tolerant (SAM-PINT) \citep{shi2021distilling}; 6) Segmentation using our proposed multi-level label correction (SAM-MLC) method; 7) Segmentation using UNet \citep{ronneberger2015u} with clean labels that are from expertise annotation (Seg-Noise Free). 
Except for SEPL (no training required) and SAM-NLC (adopt the implementation by \citep{jiang2023segment}), the rest of the segmentation methods are trained based on UNet \citep{ronneberger2015u} structure to ensure a fair comparison. Among all methods, SAM-NLC represents the baseline of SAM-based zero-shot segmentation performance. Seg-Noise Free is the upper limit where all expertise labels are available, which is the ideal case but very challenging to obtain in medical image segmentation.
We show representative images from the three datasets in Fig.\ref{fig:retrain_sam_figure}. We observe that the segmentation results obtained from directly using Direct SAM, SEPL, ADELE, and SAM-NLC are not satisfactory, as clusters of tissue boundary are misclassified as background when the boundary in the original image is blurry (red triangles). Moreover, the SEPL and ADELE tend to misclassify background regions into tissue regions (gold dash circle). Although the SAM-PINT could improve the performance slightly, it also has the issue of mistakenly classify background to foreground (gold dash circle). 
In contrast, our method (i.e., SAM-MLC) shows robustness in prediction results among the three datasets. The overall prediction performance is comparable with models trained from noise-free labels, and the segmentation results are consistent with the ground truth. Our method outperforms SAM-PINT because we consider multi-weight simultaneously, and we have a label correction module to enrich high-quality labels in the training dataset. We observe that the SEPL and ADELE demonstrate the worst performance for the BUSI dataset. We consider the SEPL and ADELE to rely on the quality of Direct SAM annotations. The SEPL and ADELE may produce a negative impact if the Direct SAM annotation has poor quality. In contrast, our method is capable of generating satisfactory results when the Direct SAM labels are less optimal.

A similar observation is found in Table. \ref{tab:retrain_sam}. We use both the Dice coefficient and accuracy to evaluate the segmentation performance. Overall, SAM-MLC brings the largest improvement from Direct SAM. 
In particular, in the CT dataset, the performance of SAM-MLC is only 2.79$\%$ (Acc for tissue) and 1.26$\%$ (Dice for tissue) lower than the ideal case when segmentation is trained from noise-free labels. Noticeably, the improvements in the X-ray (JSRT) dataset and ultrasound (BUSI) are lower than the CT dataset. This is because the SNR is generally higher in CT (Computed Tomography) compared to conventional X-ray images and ultrasound images. Therefore, the boundary in X-ray and ultrasound images are blurrier than that in CT images, making it harder to correct pseudo-labels to expertise level. Generally, the segmentation performances are the lower in the BUSI dataset (around 61$\%$ Dice score for Direct SAM prediction of tissue and around 77$\%$ Dice score for Seg-Noise Free prediction of tissue) than the performance in other two datasets. This is consistent with results presented in \citep{He2023ComputerVisionBS}. We believe that the under-performance of the BUSI dataset stems from two key factors: the lower SNR the ultrasound images and the inadequate quality of the ground truth data, where the annotator had challenges in accurately delineating boundaries of tumor region.

\begin{table*}[ht!]
\footnotesize
\centering
\caption{Evaluation metrics of our proposed model on the JSRT, CT, and BUSI dataset after removing (w/o) each module from our model under SAM noisy label.}
\begin{tabular}{cccccccccccc}
\hline
 &
   &
  \multicolumn{2}{c}{SAM-MLC (ours)} &
  \multicolumn{2}{c}{(1) w/o pixel weights} &
  \multicolumn{2}{c}{(2) w/o image weights} &
  \multicolumn{2}{c}{(3) w/o retraining} \\
                      &      & Acc    & Dice    & Acc    & Dice    & Acc    & Dice    & Acc    & Dice   \\ \hline
\multirow{2}{*}{JSRT} & Lung & 85.51\% & 91.88\% & 81.42\% & 89.40\% & 82.51\% & 90.13\% & 83.69\% & 90.78\% \\
                      & Background & 92.65\% & 96.80\% & 90.64\% & 95.99\% & 91.17\% & 96.20\% & 91.76\% & 96.44\% \\ \hline
\multirow{2}{*}{CT}   & Lung & 92.70\% & 94.67\% & 90.71\% & 93.37\% & 91.90\% & 94.39\% & 90.66\% & 93.78\% \\
                      & Background & 95.98\% & 98.55\% & 94.87\% & 98.23\% & 95.62\% & 98.47\% & 95.05\% & 98.35\% \\ \hline
\multirow{2}{*}{BUSI}   & Tumor & 66.04\%   & 68.26\%    & 64.98\%      &    66.86\%    &  64.67\%     & 67.09\%    &  64.51\%      & 67.46\%        \\
                      & Background & 81.42\%    &  96.09\%     & 80.91\%      &   96.05\%    &   80.65\%    & 95.92\%    & 80.63\%      &  95.97\%    \\ \hline
\end{tabular}
\label{tab:samablation_ct}
\end{table*}

\begin{figure*}[t]
\begin{center}
\includegraphics[width=1\textwidth]{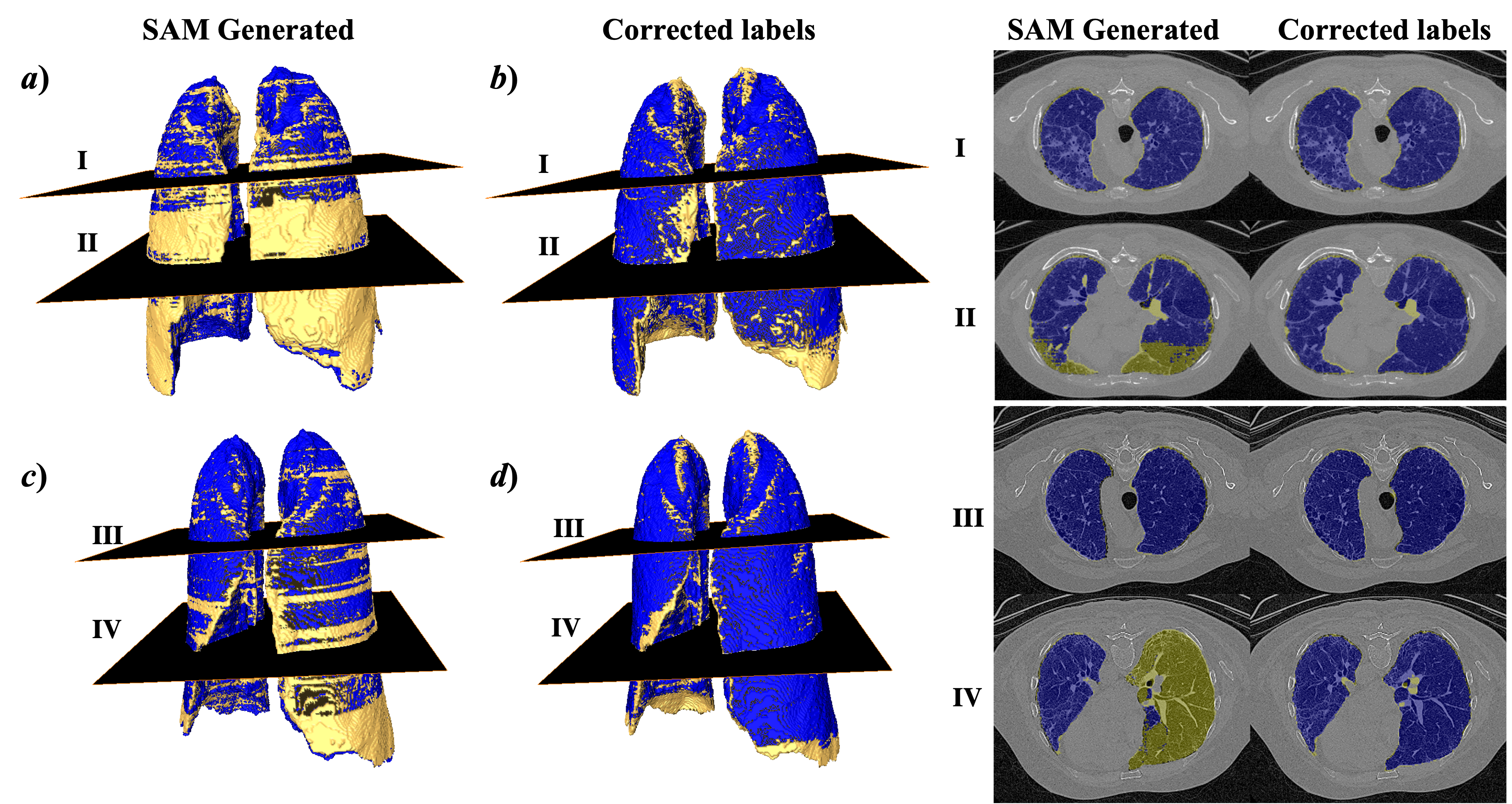}
\end{center}
\caption{Representative label correction performance in 3D. The SAM generated pseudo label is severely low quality in a) and corrected in b). The SAM generated pseudo label is mildly low quality in c) and corrected in d). Four cross-sectional 2D images from 3D visualization are shown in the right two panels. The labels are overlaid with original CT images. Yellow regions in both 3D and 2D visualization correspond to noisy pixels. Blue regions in both 3D and 2D visualization correspond to clean pixels.} 
\label{fig:3D}
\end{figure*}

\subsection{Effects of Hyperparameters}

% \textcolor{red}{stable performance within a range}

We also perform studies to evaluate the effects of two critical parameters, forget rate ($\beta$) and dropout number ($N$), on testing data, as shown in Fig. \ref{parameter}. Forget rate is a critical factor in reweighting module, determining the ratio of reliable samples used for reweighting. The dropout number is a critical factor in the self-correction module. We studied the performance with forget rate from 5\% to 90\%, and with dropout number from 60 to 140 on three datasets. 
We found that for three datasets, the performances are stable with the forget rate in the range of 5\% to 40\% and with the dropout number in the range of 70 to 110, which demonstrates that the performance is robust against the selection of hyper-parameters among all dataset.

% We found that the X-ray dataset has a lower optimal forget rate (10\% ) than the CT dataset (20\%), indicating that more samples need to be retained in X-ray to achieve effective training for segmentation. The optimal dropout number in lung CT data (120) is higher than in JSRT data (100), indicating that more dropout units are needed to quantify the uncertainty in CT dataset where the boundary of tissue region is less blurry. Overall, we found that the optimal selection of hyperparameters varies among dataset.

\subsection{Ablation Study}
To show the effectiveness of our proposed model, we further evaluate each module via ablation experiments on three datasets presenting the following four experiments, each removing a certain component of our model: (1) without the pixel-level reweighting strategy; (2) without the image-level reweighting strategy; (3) without the retraining module. The performance degenerates when any one of our modules is removed, showing the contribution of each module. Different from previous research, our reweighting strategy applies at both the image level and the pixel level. As shown in Table \ref{tab:samablation_ct}, particularly in cases (1) and (2), this combined strategy can hugely improve the segmentation performance. The results from case (1) verify that our pixel-level reweighting strategy can effectively differentiate the mis-labeled pixels from the pixels with accurate annotations. The results from case (2) show that our network successfully avoids performance degeneration by ignoring severely corrupted samples via the image-level reweighting scheme. Moreover, in case (3), the accuracy drops significantly when the retraining module is removed. These results overall confirm the beneficial effect of the label correction module and the retraining module, indicating that the label correction module can effectively enrich the clean training set labels via updating noisy labels.

\subsection{3D Visualization}

Figure \ref{fig:3D} shows the 3D visualization of label correction from the CT dataset. We compare the distribution of noisy pixels (colored in yellow) and clean pixels (colored in blue) before and after label correction in 3D space. Cross-sectional image is processed sequentially in 2D and then aligned in 3D. Figure \ref{fig:3D} (a-b) shows a scenario where SAM generated labels are generally low quality around surface. From plane I, we observe that even the boundary region is noisy, the pixels in the inner region could be dominantly clean. Figure \ref{fig:3D} (c-d) shows a scenario that is comparatively mild case of noise in 3D surface area. In plane IV, although the surface is not clustered with noisy image, the inner region (i.e., the right lobe) is severely noised. In such case, simply learning at pixel-level may learn incorrect boundary patterns. There is a need for learning at image-level such that labels that are severely corrupted can be minimally learned in training process. Overall, the pattern of consecutive corrupted frames highlights a multi-level reweight/learning strategy to consider label noise at image-level. 

Four planes at the right two columns represent scenarios where noise pixels at mild (plane I and III), moderate (plane II), and severe (plane IV). In all scenarios,  our label correction method successfully corrects a majority of noisy pixels and updates them to clean pixels.  The corrected labels accurately delineate the morphological changes in the lung region tissues. The 3D segmentation visualization indicates great potential to evaluate the 3D morphology of lung regions and calculate metrics related to respiratory disease.

%% file: sections/Discussion.tex
\section{Discussion}
%\textcolor{red}{YG: Need a dedicated section for discussion. Which will highlight our contribution, potential drawback and future work. }
% More items: 

% 1) why we only compare with one method, PINT, others require clean subset of labels. 
% 2) our computational load is low. One iteration of correction, no additional cost in testing.
% 3) Generic in other segmentation applications

In this study, we present a novel pseudo label correction framework to enhance the performance of SAM segmentation in medical images. This framework addresses the issue of low-quality pseudo labels in supervised learning and also takes advantage of zero-shot learning, in which no pixel-wised annotation is needed. Through multi-level reweighting and self-correction, we achieve significant improvements in segmentation performance compared to Direct SAM, effectively approaching the quality of fully supervised learning with expert-level annotations.

In the evaluation of label correction methods for medical segmentation, we specifically compare our framework with the SEPL \citep{Chen2023SegmentAM}, ADELE \citep{liu2022adaptive}, and PINT \citep{shi2021distilling}. Unlike the SEPL and ADELE, our method updates the labels based on the calculation of label confidence during training. Our framework is different from PINT in three aspects. First, we handle low-weight pixels differently. Unlike PINT, which discards such pixels during training for robust learning, our approach corrects and updates these pixels with new labels. This augmentation of the training dataset with additional labels contributes to improved performance. Second, our framework includes an extra retraining module that refines the network through multiple runs with updated labels, in contrast to PINT's single network training. Third, while PINT employs a two-level sequential reweighting strategy, our approach conducts reweighting concurrently. PINT employs distinct phases of pixel-level and image-level noise-tolerant learning, whereas our method ensures computational efficiency and mitigates the accumulation of low-quality image-level errors between phases. 

We did not compare it with other existing label correction methods such as \citep{zhu2019pick,mirikharaji2019learning, han2018co}, including our previous work in \citep{huang2021coseg}. The reason is that those methods require a subset of images with predefined clean labels, which is not feasible to obtain with SAM-generated pseudo labels.

Noticeably, our enhanced performance incurs only minimal addition in computational cost. A single iteration of retraining is added, and there are no additional costs during online testing. This favorable characteristic makes our framework suitable for real-time diagnosis applications.

Furthermore, we demonstrate the generality and effectiveness of our framework on three medical datasets: an X-ray dataset, a lung CT dataset, and an ultrasound dataset. The results showcase the capability of our approach in correcting pseudo labels for lung structure as well as tumor tissue. In the future, we plan to validate our framework on more intricate tissue structures, such as airway structures and irregular cancer structures, further expanding its applicability and impact in medical image segmentation.

%% file: sections/conclusion.tex
\section{Conclusion}
In this paper, we introduce a novel label correction learning framework designed to advance the boundaries of SAM-based medical segmentation. The proposed method jointly optimizes the network and the noisy training set. With special consideration on the pixel-wise and image-wise label quality, we apply a multi-level reweighting strategy for noise resistance and overfitting control. To improve the quality of labels, the self-correction module automatically corrects the noisy labels to enrich the clean pixels for further training. 

Notably, the strength of our approach lies in its ability to enhance segmentation accuracy without relying on pixel-wise annotations. We validate the effectiveness of our label correction framework through comprehensive experiments on the JSRT X-ray dataset, lung CT dataset, and Ultrasound breast cancer dataset, demonstrating its capability to significantly improve segmentation results in medical image processing. %but also is compatible with other biomedical image segmentation networks to effectively increase the noise immunity.

%% file: medima-template.bbl
\begin{thebibliography}{59}
\expandafter\ifx\csname natexlab\endcsname\relax\def\natexlab#1{#1}\fi
\providecommand{\url}[1]{\texttt{#1}}
\providecommand{\href}[2]{#2}
\providecommand{\path}[1]{#1}
\providecommand{\DOIprefix}{doi:}
\providecommand{\ArXivprefix}{arXiv:}
\providecommand{\URLprefix}{URL: }
\providecommand{\Pubmedprefix}{pmid:}
\providecommand{\doi}[1]{\href{http://dx.doi.org/#1}{\path{#1}}}
\providecommand{\Pubmed}[1]{\href{pmid:#1}{\path{#1}}}
\providecommand{\bibinfo}[2]{#2}
\ifx\xfnm\relax \def\xfnm[#1]{\unskip,\space#1}\fi
%Type = Article
\bibitem[{Al-Dhabyani et~al.(2020)Al-Dhabyani, Gomaa, Khaled and Fahmy}]{al2020dataset}
\bibinfo{author}{Al-Dhabyani, W.}, \bibinfo{author}{Gomaa, M.}, \bibinfo{author}{Khaled, H.}, \bibinfo{author}{Fahmy, A.}, \bibinfo{year}{2020}.
\newblock \bibinfo{title}{Dataset of breast ultrasound images}.
\newblock \bibinfo{journal}{Data in brief} \bibinfo{volume}{28}, \bibinfo{pages}{104863}.
%Type = Article
\bibitem[{Chen et~al.(2023)Chen, Mai, Li and Chao}]{Chen2023SegmentAM}
\bibinfo{author}{Chen, T.}, \bibinfo{author}{Mai, Z.}, \bibinfo{author}{Li, R.}, \bibinfo{author}{Chao, W.L.}, \bibinfo{year}{2023}.
\newblock \bibinfo{title}{Segment anything model (sam) enhanced pseudo labels for weakly supervised semantic segmentation}.
\newblock \bibinfo{journal}{ArXiv} \bibinfo{volume}{abs/2305.05803}.
\newblock \URLprefix \url{https://api.semanticscholar.org/CorpusID:258587943}.
%Type = Article
\bibitem[{Cheng et~al.(2023)Cheng, Qin, Jiang, Zhang, Lao and Li}]{cheng2023sam}
\bibinfo{author}{Cheng, D.}, \bibinfo{author}{Qin, Z.}, \bibinfo{author}{Jiang, Z.}, \bibinfo{author}{Zhang, S.}, \bibinfo{author}{Lao, Q.}, \bibinfo{author}{Li, K.}, \bibinfo{year}{2023}.
\newblock \bibinfo{title}{{SAM} on medical images: A comprehensive study on three prompt modes}.
\newblock \bibinfo{journal}{arXiv preprint arXiv:2305.00035} .
%Type = Article
\bibitem[{Cui et~al.(2023)Cui, Deng, Liu, Yao, Bao, Remedios, Tang and Huo}]{cui2023all}
\bibinfo{author}{Cui, C.}, \bibinfo{author}{Deng, R.}, \bibinfo{author}{Liu, Q.}, \bibinfo{author}{Yao, T.}, \bibinfo{author}{Bao, S.}, \bibinfo{author}{Remedios, L.W.}, \bibinfo{author}{Tang, Y.}, \bibinfo{author}{Huo, Y.}, \bibinfo{year}{2023}.
\newblock \bibinfo{title}{All-in-{SAM}: from weak annotation to pixel-wise nuclei segmentation with prompt-based finetuning}.
\newblock \bibinfo{journal}{arXiv preprint arXiv:2307.00290} .
%Type = Article
\bibitem[{Deng et~al.(2023)Deng, Cui, Liu, Yao, Remedios, Bao, Landman, Wheless, Coburn, Wilson et~al.}]{deng2023segment}
\bibinfo{author}{Deng, R.}, \bibinfo{author}{Cui, C.}, \bibinfo{author}{Liu, Q.}, \bibinfo{author}{Yao, T.}, \bibinfo{author}{Remedios, L.W.}, \bibinfo{author}{Bao, S.}, \bibinfo{author}{Landman, B.A.}, \bibinfo{author}{Wheless, L.E.}, \bibinfo{author}{Coburn, L.A.}, \bibinfo{author}{Wilson, K.T.}, et~al., \bibinfo{year}{2023}.
\newblock \bibinfo{title}{Segment anything model ({SAM}) for digital pathology: Assess zero-shot segmentation on whole slide imaging}.
\newblock \bibinfo{journal}{arXiv preprint arXiv:2304.04155} .
%Type = Article
\bibitem[{Dosovitskiy et~al.(2020)Dosovitskiy, Beyer, Kolesnikov, Weissenborn, Zhai, Unterthiner, Dehghani, Minderer, Heigold, Gelly et~al.}]{dosovitskiy2020image}
\bibinfo{author}{Dosovitskiy, A.}, \bibinfo{author}{Beyer, L.}, \bibinfo{author}{Kolesnikov, A.}, \bibinfo{author}{Weissenborn, D.}, \bibinfo{author}{Zhai, X.}, \bibinfo{author}{Unterthiner, T.}, \bibinfo{author}{Dehghani, M.}, \bibinfo{author}{Minderer, M.}, \bibinfo{author}{Heigold, G.}, \bibinfo{author}{Gelly, S.}, et~al., \bibinfo{year}{2020}.
\newblock \bibinfo{title}{An image is worth 16x16 words: Transformers for image recognition at scale}.
\newblock \bibinfo{journal}{arXiv preprint arXiv:2010.11929} .
%Type = Article
\bibitem[{Feng et~al.(2023)Feng, Wang, Li, Ji and Liu}]{feng2023weakly}
\bibinfo{author}{Feng, J.}, \bibinfo{author}{Wang, X.}, \bibinfo{author}{Li, T.}, \bibinfo{author}{Ji, S.}, \bibinfo{author}{Liu, W.}, \bibinfo{year}{2023}.
\newblock \bibinfo{title}{Weakly-supervised semantic segmentation via online pseudo-mask correcting}.
\newblock \bibinfo{journal}{Pattern Recognition Letters} \bibinfo{volume}{165}, \bibinfo{pages}{33--38}.
%Type = Inproceedings
\bibitem[{Gal and Ghahramani(2016)}]{gal2016}
\bibinfo{author}{Gal, Y.}, \bibinfo{author}{Ghahramani, Z.}, \bibinfo{year}{2016}.
\newblock \bibinfo{title}{Dropout as a bayesian approximation: Representing model uncertainty in deep learning}, in: \bibinfo{booktitle}{International Conference on Machine Learning}, pp. \bibinfo{pages}{1050--1059}.
%Type = Inproceedings
\bibitem[{Goldberger and Ben-Reuven(2017)}]{goldberger2016training}
\bibinfo{author}{Goldberger, J.}, \bibinfo{author}{Ben-Reuven, E.}, \bibinfo{year}{2017}.
\newblock \bibinfo{title}{Training deep neural-networks using a noise adaptation layer}, in: \bibinfo{booktitle}{ICLR}.
%Type = Inproceedings
\bibitem[{Han et~al.(2018)Han, Yao, Yu, Niu, Xu, Hu, Tsang and Sugiyama}]{han2018co}
\bibinfo{author}{Han, B.}, \bibinfo{author}{Yao, Q.}, \bibinfo{author}{Yu, X.}, \bibinfo{author}{Niu, G.}, \bibinfo{author}{Xu, M.}, \bibinfo{author}{Hu, W.}, \bibinfo{author}{Tsang, I.}, \bibinfo{author}{Sugiyama, M.}, \bibinfo{year}{2018}.
\newblock \bibinfo{title}{Co-teaching: Robust training of deep neural networks with extremely noisy labels}, in: \bibinfo{booktitle}{Advances in Neural Information Processing Systems}, pp. \bibinfo{pages}{8527--8537}.
%Type = Article
\bibitem[{He et~al.(2023a)He, Bao, Li, Grant and Ou}]{he2023accuracy}
\bibinfo{author}{He, S.}, \bibinfo{author}{Bao, R.}, \bibinfo{author}{Li, J.}, \bibinfo{author}{Grant, P.E.}, \bibinfo{author}{Ou, Y.}, \bibinfo{year}{2023}a.
\newblock \bibinfo{title}{Accuracy of segment-anything model ({SAM}) in medical image segmentation tasks}.
\newblock \bibinfo{journal}{arXiv preprint arXiv:2304.09324} .
%Type = Inproceedings
\bibitem[{He et~al.(2023b)He, Bao, Li, Stout, Bj{\o}rnerud, Grant and Ou}]{He2023ComputerVisionBS}
\bibinfo{author}{He, S.}, \bibinfo{author}{Bao, R.}, \bibinfo{author}{Li, J.}, \bibinfo{author}{Stout, J.N.}, \bibinfo{author}{Bj{\o}rnerud, A.}, \bibinfo{author}{Grant, P.E.}, \bibinfo{author}{Ou, Y.}, \bibinfo{year}{2023}b.
\newblock \bibinfo{title}{Computer-vision benchmark segment-anything model (sam) in medical images: Accuracy in 12 datasets}.
\newblock \URLprefix \url{https://api.semanticscholar.org/CorpusID:258558130}.
%Type = Inproceedings
\bibitem[{He et~al.(2019)He, Yang, Li, Li, Chang and Yu}]{he2019non}
\bibinfo{author}{He, X.}, \bibinfo{author}{Yang, S.}, \bibinfo{author}{Li, G.}, \bibinfo{author}{Li, H.}, \bibinfo{author}{Chang, H.}, \bibinfo{author}{Yu, Y.}, \bibinfo{year}{2019}.
\newblock \bibinfo{title}{Non-local context encoder: Robust biomedical image segmentation against adversarial attacks}, in: \bibinfo{booktitle}{Proceedings of the AAAI Conference on Artificial Intelligence}, pp. \bibinfo{pages}{8417--8424}.
%Type = Inproceedings
\bibitem[{Hendrycks et~al.(2018)Hendrycks, Mazeika, Wilson and Gimpel}]{NEURIPS2018_ad554d8c}
\bibinfo{author}{Hendrycks, D.}, \bibinfo{author}{Mazeika, M.}, \bibinfo{author}{Wilson, D.}, \bibinfo{author}{Gimpel, K.}, \bibinfo{year}{2018}.
\newblock \bibinfo{title}{Using trusted data to train deep networks on labels corrupted by severe noise}, in: \bibinfo{editor}{Bengio, S.}, \bibinfo{editor}{Wallach, H.}, \bibinfo{editor}{Larochelle, H.}, \bibinfo{editor}{Grauman, K.}, \bibinfo{editor}{Cesa-Bianchi, N.}, \bibinfo{editor}{Garnett, R.} (Eds.), \bibinfo{booktitle}{Advances in Neural Information Processing Systems}, \bibinfo{publisher}{Curran Associates, Inc.}
%Type = Article
\bibitem[{Hu and Li(2023)}]{hu2023sam}
\bibinfo{author}{Hu, C.}, \bibinfo{author}{Li, X.}, \bibinfo{year}{2023}.
\newblock \bibinfo{title}{When {SAM} meets medical images: An investigation of segment anything model ({SAM}) on multi-phase liver tumor segmentation}.
\newblock \bibinfo{journal}{arXiv preprint arXiv:2304.08506} .
%Type = Inproceedings
\bibitem[{Hu et~al.(2019)Hu, Worrall, Knegt, Veeling, Huisman and Welling}]{hu2019supervised}
\bibinfo{author}{Hu, S.}, \bibinfo{author}{Worrall, D.}, \bibinfo{author}{Knegt, S.}, \bibinfo{author}{Veeling, B.}, \bibinfo{author}{Huisman, H.}, \bibinfo{author}{Welling, M.}, \bibinfo{year}{2019}.
\newblock \bibinfo{title}{Supervised uncertainty quantification for segmentation with multiple annotations}, in: \bibinfo{booktitle}{International Conference on Medical Image Computing and Computer-Assisted Intervention}, \bibinfo{organization}{Springer}. pp. \bibinfo{pages}{137--145}.
%Type = Misc
\bibitem[{Huang et~al.(2021)Huang, Zhang, Laine, Angelini, Hendon and Gan}]{huang2021coseg}
\bibinfo{author}{Huang, Z.}, \bibinfo{author}{Zhang, H.}, \bibinfo{author}{Laine, A.}, \bibinfo{author}{Angelini, E.}, \bibinfo{author}{Hendon, C.}, \bibinfo{author}{Gan, Y.}, \bibinfo{year}{2021}.
\newblock \bibinfo{title}{Co-seg: An image segmentation framework against label corruption}.
\newblock \URLprefix \url{https://arxiv.org/pdf/2102.00523.pdf}, \href{http://arxiv.org/abs/2102.00523}{\tt arXiv:2102.00523}.
%Type = Incollection
\bibitem[{Hwang and Park(2017)}]{hwang2017accurate}
\bibinfo{author}{Hwang, S.}, \bibinfo{author}{Park, S.}, \bibinfo{year}{2017}.
\newblock \bibinfo{title}{Accurate lung segmentation via network-wise training of convolutional networks}, in: \bibinfo{booktitle}{Deep Learning in Medical Image Analysis and Multimodal Learning for Clinical Decision Support}. \bibinfo{publisher}{Springer}, pp. \bibinfo{pages}{92--99}.
%Type = Inproceedings
\bibitem[{Ibrahim et~al.(2020)Ibrahim, Vahdat, Ranjbar and Macready}]{ibrahim2020semi}
\bibinfo{author}{Ibrahim, M.S.}, \bibinfo{author}{Vahdat, A.}, \bibinfo{author}{Ranjbar, M.}, \bibinfo{author}{Macready, W.G.}, \bibinfo{year}{2020}.
\newblock \bibinfo{title}{Semi-supervised semantic image segmentation with self-correcting networks}, in: \bibinfo{booktitle}{Proceedings of the IEEE/CVF conference on computer vision and pattern recognition}, pp. \bibinfo{pages}{12715--12725}.
%Type = Article
\bibitem[{Ji et~al.(2023a)Ji, Fan, Xu, Cheng, Zhou and Van~Gool}]{ji2023sam}
\bibinfo{author}{Ji, G.P.}, \bibinfo{author}{Fan, D.P.}, \bibinfo{author}{Xu, P.}, \bibinfo{author}{Cheng, M.M.}, \bibinfo{author}{Zhou, B.}, \bibinfo{author}{Van~Gool, L.}, \bibinfo{year}{2023}a.
\newblock \bibinfo{title}{{SAM} struggles in concealed scenes--empirical study on ``segment anything"}.
\newblock \bibinfo{journal}{arXiv preprint arXiv:2304.06022} .
%Type = Article
\bibitem[{Ji et~al.(2023b)Ji, Li, Bi, Li and Cheng}]{ji2023segment}
\bibinfo{author}{Ji, W.}, \bibinfo{author}{Li, J.}, \bibinfo{author}{Bi, Q.}, \bibinfo{author}{Li, W.}, \bibinfo{author}{Cheng, L.}, \bibinfo{year}{2023}b.
\newblock \bibinfo{title}{Segment anything is not always perfect: An investigation of {SAM} on different real-world applications}.
\newblock \bibinfo{journal}{arXiv preprint arXiv:2304.05750} .
%Type = Article
\bibitem[{Jiang and Yang(2023)}]{jiang2023segment}
\bibinfo{author}{Jiang, P.T.}, \bibinfo{author}{Yang, Y.}, \bibinfo{year}{2023}.
\newblock \bibinfo{title}{Segment anything is a good pseudo-label generator for weakly supervised semantic segmentation}.
\newblock \bibinfo{journal}{arXiv preprint arXiv:2305.01275} .
%Type = Inproceedings
\bibitem[{Jindal et~al.(2016)Jindal, Nokleby and Chen}]{jindal2016learning}
\bibinfo{author}{Jindal, I.}, \bibinfo{author}{Nokleby, M.}, \bibinfo{author}{Chen, X.}, \bibinfo{year}{2016}.
\newblock \bibinfo{title}{Learning deep networks from noisy labels with dropout regularization}, in: \bibinfo{booktitle}{2016 IEEE 16th International Conference on Data Mining (ICDM)}, \bibinfo{organization}{IEEE}. pp. \bibinfo{pages}{967--972}.
%Type = Inproceedings
\bibitem[{Kalapos and Gyires-T{\'o}th(2023)}]{kalapos2023self}
\bibinfo{author}{Kalapos, A.}, \bibinfo{author}{Gyires-T{\'o}th, B.}, \bibinfo{year}{2023}.
\newblock \bibinfo{title}{Self-supervised pretraining for {2D} medical image segmentation}, in: \bibinfo{booktitle}{Computer Vision--ECCV 2022 Workshops: Tel Aviv, Israel, October 23--27, 2022, Proceedings, Part VII}, \bibinfo{organization}{Springer}. pp. \bibinfo{pages}{472--484}.
%Type = Article
\bibitem[{Kendall et~al.(2015)Kendall, Badrinarayanan and Cipolla}]{kendall2015bayesian}
\bibinfo{author}{Kendall, A.}, \bibinfo{author}{Badrinarayanan, V.}, \bibinfo{author}{Cipolla, R.}, \bibinfo{year}{2015}.
\newblock \bibinfo{title}{Bayesian segnet: Model uncertainty in deep convolutional encoder-decoder architectures for scene understanding}.
\newblock \bibinfo{journal}{arXiv preprint arXiv:1511.02680} .
%Type = Article
\bibitem[{Kingma and Ba(2014)}]{kingma2014adam}
\bibinfo{author}{Kingma, D.P.}, \bibinfo{author}{Ba, J.}, \bibinfo{year}{2014}.
\newblock \bibinfo{title}{Adam: A method for stochastic optimization}.
\newblock \bibinfo{journal}{arXiv preprint arXiv:1412.6980} .
%Type = Article
\bibitem[{Kirillov et~al.(2023)Kirillov, Mintun, Ravi, Mao, Rolland, Gustafson, Xiao, Whitehead, Berg, Lo et~al.}]{kirillov2023segment}
\bibinfo{author}{Kirillov, A.}, \bibinfo{author}{Mintun, E.}, \bibinfo{author}{Ravi, N.}, \bibinfo{author}{Mao, H.}, \bibinfo{author}{Rolland, C.}, \bibinfo{author}{Gustafson, L.}, \bibinfo{author}{Xiao, T.}, \bibinfo{author}{Whitehead, S.}, \bibinfo{author}{Berg, A.C.}, \bibinfo{author}{Lo, W.Y.}, et~al., \bibinfo{year}{2023}.
\newblock \bibinfo{title}{Segment anything}.
\newblock \bibinfo{journal}{arXiv preprint arXiv:2304.02643} .
%Type = Misc
\bibitem[{{Konya}(2020)}]{lungCT2020}
\bibinfo{author}{{Konya}}, \bibinfo{year}{2020}.
\newblock \bibinfo{title}{Lung segmentation dataset}.
\newblock \URLprefix \url{https://www.kaggle.com/sandorkonya/ct-lung-heart-trachea-segmentation}.
%Type = Article
\bibitem[{Li et~al.(2023)Li, Xiong, Qiu, Pan, Luo and Zhang}]{li2023segment}
\bibinfo{author}{Li, N.}, \bibinfo{author}{Xiong, L.}, \bibinfo{author}{Qiu, W.}, \bibinfo{author}{Pan, Y.}, \bibinfo{author}{Luo, Y.}, \bibinfo{author}{Zhang, Y.}, \bibinfo{year}{2023}.
\newblock \bibinfo{title}{Segment anything model for semi-supervised medical image segmentation via selecting reliable pseudo-labels}.
\newblock \bibinfo{journal}{Available at SSRN 4477443} .
%Type = Article
\bibitem[{Li et~al.(2022)Li, Yang, Shu, Yu, Huang, Li, Chen, Hu, Shu, Yu et~al.}]{li2022research}
\bibinfo{author}{Li, Z.}, \bibinfo{author}{Yang, L.}, \bibinfo{author}{Shu, L.}, \bibinfo{author}{Yu, Z.}, \bibinfo{author}{Huang, J.}, \bibinfo{author}{Li, J.}, \bibinfo{author}{Chen, L.}, \bibinfo{author}{Hu, S.}, \bibinfo{author}{Shu, T.}, \bibinfo{author}{Yu, G.}, et~al., \bibinfo{year}{2022}.
\newblock \bibinfo{title}{Research on {CT} lung segmentation method of preschool children based on traditional image processing and resunet}.
\newblock \bibinfo{journal}{Computational and Mathematical Methods in Medicine} \bibinfo{volume}{2022}.
%Type = Inproceedings
\bibitem[{Liu et~al.(2022)Liu, Liu, Zhu, Shen and Fernandez–Granda}]{liu2022adaptive}
\bibinfo{author}{Liu, S.}, \bibinfo{author}{Liu, K.}, \bibinfo{author}{Zhu, W.}, \bibinfo{author}{Shen, Y.}, \bibinfo{author}{Fernandez–Granda, C.}, \bibinfo{year}{2022}.
\newblock \bibinfo{title}{Adaptive early-learning correction for segmentation from noisy annotations}, in: \bibinfo{booktitle}{2022 IEEE/CVF Conference on Computer Vision and Pattern Recognition (CVPR)}, pp. \bibinfo{pages}{2596--2606}.
%Type = Incollection
\bibitem[{Mirikharaji et~al.(2019)Mirikharaji, Yan and Hamarneh}]{mirikharaji2019learning}
\bibinfo{author}{Mirikharaji, Z.}, \bibinfo{author}{Yan, Y.}, \bibinfo{author}{Hamarneh, G.}, \bibinfo{year}{2019}.
\newblock \bibinfo{title}{Learning to segment skin lesions from noisy annotations}, in: \bibinfo{booktitle}{Domain Adaptation and Representation Transfer and Medical Image Learning with Less Labels and Imperfect Data}. \bibinfo{publisher}{Springer}, pp. \bibinfo{pages}{207--215}.
%Type = Article
\bibitem[{Mo and Tian(2023)}]{mo2023av}
\bibinfo{author}{Mo, S.}, \bibinfo{author}{Tian, Y.}, \bibinfo{year}{2023}.
\newblock \bibinfo{title}{Av-sam: Segment anything model meets audio-visual localization and segmentation}.
\newblock \bibinfo{journal}{arXiv preprint arXiv:2305.01836} .
%Type = Article
\bibitem[{Ouyang et~al.(2022)Ouyang, Biffi, Chen, Kart, Qiu and Rueckert}]{ouyang2022self}
\bibinfo{author}{Ouyang, C.}, \bibinfo{author}{Biffi, C.}, \bibinfo{author}{Chen, C.}, \bibinfo{author}{Kart, T.}, \bibinfo{author}{Qiu, H.}, \bibinfo{author}{Rueckert, D.}, \bibinfo{year}{2022}.
\newblock \bibinfo{title}{Self-supervised learning for few-shot medical image segmentation}.
\newblock \bibinfo{journal}{IEEE Transactions on Medical Imaging} \bibinfo{volume}{41}, \bibinfo{pages}{1837--1848}.
%Type = Inproceedings
\bibitem[{Patrini et~al.(2017)Patrini, Rozza, Krishna~Menon, Nock and Qu}]{patrini2017making}
\bibinfo{author}{Patrini, G.}, \bibinfo{author}{Rozza, A.}, \bibinfo{author}{Krishna~Menon, A.}, \bibinfo{author}{Nock, R.}, \bibinfo{author}{Qu, L.}, \bibinfo{year}{2017}.
\newblock \bibinfo{title}{Making deep neural networks robust to label noise: A loss correction approach}, in: \bibinfo{booktitle}{Proceedings of the IEEE Conference on Computer Vision and Pattern Recognition}, pp. \bibinfo{pages}{1944--1952}.
%Type = Inproceedings
\bibitem[{Quan et~al.(2020)Quan, Li, Chen and Zhang}]{quan2020effective}
\bibinfo{author}{Quan, L.}, \bibinfo{author}{Li, Y.}, \bibinfo{author}{Chen, X.}, \bibinfo{author}{Zhang, N.}, \bibinfo{year}{2020}.
\newblock \bibinfo{title}{An effective data refinement approach for upper gastrointestinal anatomy recognition}, in: \bibinfo{booktitle}{International Conference on Medical Image Computing and Computer-Assisted Intervention}, \bibinfo{organization}{Springer}. pp. \bibinfo{pages}{43--52}.
%Type = Inproceedings
\bibitem[{Ronneberger et~al.(2015)Ronneberger, Fischer and Brox}]{ronneberger2015u}
\bibinfo{author}{Ronneberger, O.}, \bibinfo{author}{Fischer, P.}, \bibinfo{author}{Brox, T.}, \bibinfo{year}{2015}.
\newblock \bibinfo{title}{{U-Net}: Convolutional networks for biomedical image segmentation}, in: \bibinfo{booktitle}{International Conference on Medical image computing and computer-assisted intervention}, \bibinfo{organization}{Springer}. pp. \bibinfo{pages}{234--241}.
%Type = Article
\bibitem[{Roy et~al.(2023)Roy, Wald, Koehler, Rokuss, Disch, Holzschuh, Zimmerer and Maier-Hein}]{roy2023sam}
\bibinfo{author}{Roy, S.}, \bibinfo{author}{Wald, T.}, \bibinfo{author}{Koehler, G.}, \bibinfo{author}{Rokuss, M.R.}, \bibinfo{author}{Disch, N.}, \bibinfo{author}{Holzschuh, J.}, \bibinfo{author}{Zimmerer, D.}, \bibinfo{author}{Maier-Hein, K.H.}, \bibinfo{year}{2023}.
\newblock \bibinfo{title}{{SAM.MD}: Zero-shot medical image segmentation capabilities of the segment anything model}.
\newblock \bibinfo{journal}{arXiv preprint arXiv:2304.05396} .
%Type = Incollection
\bibitem[{Sedai et~al.(2018)Sedai, Antony, Mahapatra and Garnavi}]{sedai2018joint}
\bibinfo{author}{Sedai, S.}, \bibinfo{author}{Antony, B.}, \bibinfo{author}{Mahapatra, D.}, \bibinfo{author}{Garnavi, R.}, \bibinfo{year}{2018}.
\newblock \bibinfo{title}{Joint segmentation and uncertainty visualization of retinal layers in optical coherence tomography images using bayesian deep learning}, in: \bibinfo{booktitle}{Computational Pathology and Ophthalmic Medical Image Analysis}. \bibinfo{publisher}{Springer}, pp. \bibinfo{pages}{219--227}.
%Type = Article
\bibitem[{Shen et~al.(2021)Shen, Shamout, Oliver, Witowski, Kannan, Park, Wu, Huddleston, Wolfson, Millet, Ehrenpreis, Awal, Tyma, Samreen, Gao, Chhor, Gandhi, Lee, Kumari-Subaiya, Leonard, Mohammed, Moczulski, Altabet, Babb, Lewin, Reig, Moy, Heacock and Geras}]{Shen.2021}
\bibinfo{author}{Shen, Y.}, \bibinfo{author}{Shamout, F.E.}, \bibinfo{author}{Oliver, J.R.}, \bibinfo{author}{Witowski, J.}, \bibinfo{author}{Kannan, K.}, \bibinfo{author}{Park, J.}, \bibinfo{author}{Wu, N.}, \bibinfo{author}{Huddleston, C.}, \bibinfo{author}{Wolfson, S.}, \bibinfo{author}{Millet, A.}, \bibinfo{author}{Ehrenpreis, R.}, \bibinfo{author}{Awal, D.}, \bibinfo{author}{Tyma, C.}, \bibinfo{author}{Samreen, N.}, \bibinfo{author}{Gao, Y.}, \bibinfo{author}{Chhor, C.}, \bibinfo{author}{Gandhi, S.}, \bibinfo{author}{Lee, C.}, \bibinfo{author}{Kumari-Subaiya, S.}, \bibinfo{author}{Leonard, C.}, \bibinfo{author}{Mohammed, R.}, \bibinfo{author}{Moczulski, C.}, \bibinfo{author}{Altabet, J.}, \bibinfo{author}{Babb, J.}, \bibinfo{author}{Lewin, A.}, \bibinfo{author}{Reig, B.}, \bibinfo{author}{Moy, L.}, \bibinfo{author}{Heacock, L.}, \bibinfo{author}{Geras, K.J.}, \bibinfo{year}{2021}.
\newblock \bibinfo{title}{Artificial intelligence system reduces false-positive findings in the interpretation of breast ultrasound exams}.
\newblock \bibinfo{journal}{Nature Communications} \bibinfo{volume}{12}, \bibinfo{pages}{5645}.
\newblock \DOIprefix\doi{10.1038/s41467-021-26023-2}.
%Type = Inproceedings
\bibitem[{Shi and Wu(2021)}]{shi2021distilling}
\bibinfo{author}{Shi, J.}, \bibinfo{author}{Wu, J.}, \bibinfo{year}{2021}.
\newblock \bibinfo{title}{Distilling effective supervision for robust medical image segmentation with noisy labels}, in: \bibinfo{booktitle}{Medical Image Computing and Computer Assisted Intervention--MICCAI 2021: 24th International Conference, Strasbourg, France, September 27--October 1, 2021, Proceedings, Part I 24}, \bibinfo{organization}{Springer}. pp. \bibinfo{pages}{668--677}.
%Type = Article
\bibitem[{Shi et~al.(2023)Shi, Qiu, Abaxi, Wei, Lo and Yuan}]{shi2023generalist}
\bibinfo{author}{Shi, P.}, \bibinfo{author}{Qiu, J.}, \bibinfo{author}{Abaxi, S.M.D.}, \bibinfo{author}{Wei, H.}, \bibinfo{author}{Lo, F.P.W.}, \bibinfo{author}{Yuan, W.}, \bibinfo{year}{2023}.
\newblock \bibinfo{title}{Generalist vision foundation models for medical imaging: A case study of segment anything model on zero-shot medical segmentation}.
\newblock \bibinfo{journal}{Diagnostics} \bibinfo{volume}{13}, \bibinfo{pages}{1947}.
%Type = Article
\bibitem[{Shiraishi et~al.(2000)Shiraishi, Katsuragawa, Ikezoe, Matsumoto, Kobayashi, Komatsu, Matsui, Fujita, Kodera and Doi}]{shiraishi2000development}
\bibinfo{author}{Shiraishi, J.}, \bibinfo{author}{Katsuragawa, S.}, \bibinfo{author}{Ikezoe, J.}, \bibinfo{author}{Matsumoto, T.}, \bibinfo{author}{Kobayashi, T.}, \bibinfo{author}{Komatsu, K.i.}, \bibinfo{author}{Matsui, M.}, \bibinfo{author}{Fujita, H.}, \bibinfo{author}{Kodera, Y.}, \bibinfo{author}{Doi, K.}, \bibinfo{year}{2000}.
\newblock \bibinfo{title}{Development of a digital image database for chest radiographs with and without a lung nodule: receiver operating characteristic analysis of radiologists' detection of pulmonary nodules}.
\newblock \bibinfo{journal}{American Journal of Roentgenology} \bibinfo{volume}{174}, \bibinfo{pages}{71--74}.
%Type = Inproceedings
\bibitem[{Sukhbaatar et~al.(2015)Sukhbaatar, Bruna, Paluri, Bourdev and Fergus}]{sukhbaatar2015training}
\bibinfo{author}{Sukhbaatar, S.}, \bibinfo{author}{Bruna, J.}, \bibinfo{author}{Paluri, M.}, \bibinfo{author}{Bourdev, L.}, \bibinfo{author}{Fergus, R.}, \bibinfo{year}{2015}.
\newblock \bibinfo{title}{Training convolutional networks with noisy labels}, in: \bibinfo{booktitle}{3rd International Conference on Learning Representations, ICLR 2015}.
%Type = Article
\bibitem[{Sun et~al.(2023)Sun, Liu, Zhang, Zhong and Barnes}]{sun2023alternative}
\bibinfo{author}{Sun, W.}, \bibinfo{author}{Liu, Z.}, \bibinfo{author}{Zhang, Y.}, \bibinfo{author}{Zhong, Y.}, \bibinfo{author}{Barnes, N.}, \bibinfo{year}{2023}.
\newblock \bibinfo{title}{An alternative to wsss? an empirical study of the segment anything model (sam) on weakly-supervised semantic segmentation problems}.
\newblock \bibinfo{journal}{arXiv preprint arXiv:2305.01586} .
%Type = Article
\bibitem[{Van~Ginneken et~al.(2006)Van~Ginneken, Stegmann and Loog}]{van2006segmentation}
\bibinfo{author}{Van~Ginneken, B.}, \bibinfo{author}{Stegmann, M.B.}, \bibinfo{author}{Loog, M.}, \bibinfo{year}{2006}.
\newblock \bibinfo{title}{Segmentation of anatomical structures in chest radiographs using supervised methods: a comparative study on a public database}.
\newblock \bibinfo{journal}{Medical Image Analysis} \bibinfo{volume}{10}, \bibinfo{pages}{19--40}.
%Type = Article
\bibitem[{{Walid Al-Dhabyani} et~al.(2020){Walid Al-Dhabyani}, {Mohammed Gomaa}, {Hussien Khaled} and {Aly Fahmy}}]{WalidAlDhabyani.2020}
\bibinfo{author}{{Walid Al-Dhabyani}}, \bibinfo{author}{{Mohammed Gomaa}}, \bibinfo{author}{{Hussien Khaled}}, \bibinfo{author}{{Aly Fahmy}}, \bibinfo{year}{2020}.
\newblock \bibinfo{title}{Dataset of breast ultrasound images}.
\newblock \bibinfo{journal}{Data in Brief} \bibinfo{volume}{28}, \bibinfo{pages}{104863}.
\newblock \URLprefix \url{https://www.sciencedirect.com/science/article/pii/S2352340919312181}, \DOIprefix\doi{10.1016/j.dib.2019.104863}.
%Type = Article
\bibitem[{Wang et~al.(2023)Wang, Zhang, Du, Tao and Zhang}]{wang2023scaling}
\bibinfo{author}{Wang, D.}, \bibinfo{author}{Zhang, J.}, \bibinfo{author}{Du, B.}, \bibinfo{author}{Tao, D.}, \bibinfo{author}{Zhang, L.}, \bibinfo{year}{2023}.
\newblock \bibinfo{title}{Scaling-up remote sensing segmentation dataset with segment anything model}.
\newblock \bibinfo{journal}{arXiv preprint arXiv:2305.02034} .
%Type = Inproceedings
\bibitem[{Wang et~al.(2020a)Wang, Zhou, Fang, Wang and Wang}]{wang2020meta}
\bibinfo{author}{Wang, J.}, \bibinfo{author}{Zhou, S.}, \bibinfo{author}{Fang, C.}, \bibinfo{author}{Wang, L.}, \bibinfo{author}{Wang, J.}, \bibinfo{year}{2020}a.
\newblock \bibinfo{title}{Meta corrupted pixels mining for medical image segmentation}, in: \bibinfo{booktitle}{International Conference on Medical Image Computing and Computer-Assisted Intervention}, \bibinfo{organization}{Springer}. pp. \bibinfo{pages}{335--345}.
%Type = Inproceedings
\bibitem[{Wang et~al.(2020b)Wang, Hu and Hu}]{wang2020training}
\bibinfo{author}{Wang, Z.}, \bibinfo{author}{Hu, G.}, \bibinfo{author}{Hu, Q.}, \bibinfo{year}{2020}b.
\newblock \bibinfo{title}{Training noise-robust deep neural networks via meta-learning}, in: \bibinfo{booktitle}{Proceedings of the IEEE/CVF Conference on Computer Vision and Pattern Recognition}, pp. \bibinfo{pages}{4524--4533}.
%Type = Inproceedings
\bibitem[{Wu et~al.(2022)Wu, Wei, Tan and Zhao}]{wu2022pseudo}
\bibinfo{author}{Wu, H.}, \bibinfo{author}{Wei, S.}, \bibinfo{author}{Tan, C.}, \bibinfo{author}{Zhao, Y.}, \bibinfo{year}{2022}.
\newblock \bibinfo{title}{Pseudo-label correction from pixel to image}, in: \bibinfo{booktitle}{2022 4th International Conference on Advances in Computer Technology, Information Science and Communications (CTISC)}, \bibinfo{organization}{IEEE}. pp. \bibinfo{pages}{1--5}.
%Type = Article
\bibitem[{Yi et~al.(2021)Yi, Huang, Guan, Pu and Zhang}]{yi2021learning}
\bibinfo{author}{Yi, R.}, \bibinfo{author}{Huang, Y.}, \bibinfo{author}{Guan, Q.}, \bibinfo{author}{Pu, M.}, \bibinfo{author}{Zhang, R.}, \bibinfo{year}{2021}.
\newblock \bibinfo{title}{Learning from pixel-level label noise: A new perspective for semi-supervised semantic segmentation}.
\newblock \bibinfo{journal}{IEEE Transactions on Image Processing} \bibinfo{volume}{31}, \bibinfo{pages}{623--635}.
%Type = Inproceedings
\bibitem[{Zhang et~al.(2020a)Zhang, Gao, Lyu, Zhao, Wang, Ding, Wang, Li and Cui}]{zhang2020characterizing}
\bibinfo{author}{Zhang, M.}, \bibinfo{author}{Gao, J.}, \bibinfo{author}{Lyu, Z.}, \bibinfo{author}{Zhao, W.}, \bibinfo{author}{Wang, Q.}, \bibinfo{author}{Ding, W.}, \bibinfo{author}{Wang, S.}, \bibinfo{author}{Li, Z.}, \bibinfo{author}{Cui, S.}, \bibinfo{year}{2020}a.
\newblock \bibinfo{title}{Characterizing label errors: Confident learning for noisy-labeled image segmentation}, in: \bibinfo{booktitle}{International Conference on Medical Image Computing and Computer-Assisted Intervention}, \bibinfo{organization}{Springer}. pp. \bibinfo{pages}{721--730}.
%Type = Article
\bibitem[{Zhang et~al.(2023a)Zhang, Jiang, Guo, Yan, Pan, Dong, Gao and Li}]{zhang2023personalize}
\bibinfo{author}{Zhang, R.}, \bibinfo{author}{Jiang, Z.}, \bibinfo{author}{Guo, Z.}, \bibinfo{author}{Yan, S.}, \bibinfo{author}{Pan, J.}, \bibinfo{author}{Dong, H.}, \bibinfo{author}{Gao, P.}, \bibinfo{author}{Li, H.}, \bibinfo{year}{2023}a.
\newblock \bibinfo{title}{Personalize segment anything model with one shot}.
\newblock \bibinfo{journal}{arXiv preprint arXiv:2305.03048} .
%Type = Article
\bibitem[{Zhang and Jiao(2023)}]{zhang2023segment}
\bibinfo{author}{Zhang, Y.}, \bibinfo{author}{Jiao, R.}, \bibinfo{year}{2023}.
\newblock \bibinfo{title}{How segment anything model ({SAM}) boost medical image segmentation?}
\newblock \bibinfo{journal}{arXiv preprint arXiv:2305.03678} .
%Type = Article
\bibitem[{Zhang et~al.(2023b)Zhang, Zhou, Liang and Chen}]{zhang2023input}
\bibinfo{author}{Zhang, Y.}, \bibinfo{author}{Zhou, T.}, \bibinfo{author}{Liang, P.}, \bibinfo{author}{Chen, D.Z.}, \bibinfo{year}{2023}b.
\newblock \bibinfo{title}{Input augmentation with {SAM}: Boosting medical image segmentation with segmentation foundation model}.
\newblock \bibinfo{journal}{arXiv preprint arXiv:2304.11332} .
%Type = Inproceedings
\bibitem[{Zhang et~al.(2020b)Zhang, Zhang, Arik, Lee and Pfister}]{zhang2020distilling}
\bibinfo{author}{Zhang, Z.}, \bibinfo{author}{Zhang, H.}, \bibinfo{author}{Arik, S.O.}, \bibinfo{author}{Lee, H.}, \bibinfo{author}{Pfister, T.}, \bibinfo{year}{2020}b.
\newblock \bibinfo{title}{Distilling effective supervision from severe label noise}, in: \bibinfo{booktitle}{Proceedings of the IEEE/CVF Conference on Computer Vision and Pattern Recognition}, pp. \bibinfo{pages}{9294--9303}.
%Type = Article
\bibitem[{Zhu et~al.(2020)Zhu, Adeli, Shi, Shen and Initiative}]{zhu2020fcn}
\bibinfo{author}{Zhu, H.}, \bibinfo{author}{Adeli, E.}, \bibinfo{author}{Shi, F.}, \bibinfo{author}{Shen, D.}, \bibinfo{author}{Initiative, A.D.N.}, \bibinfo{year}{2020}.
\newblock \bibinfo{title}{{FCN} based label correction for multi-atlas guided organ segmentation}.
\newblock \bibinfo{journal}{Neuroinformatics} \bibinfo{volume}{18}, \bibinfo{pages}{319--331}.
%Type = Inproceedings
\bibitem[{Zhu et~al.(2019)Zhu, Shi and Wu}]{zhu2019pick}
\bibinfo{author}{Zhu, H.}, \bibinfo{author}{Shi, J.}, \bibinfo{author}{Wu, J.}, \bibinfo{year}{2019}.
\newblock \bibinfo{title}{Pick-and-learn: Automatic quality evaluation for noisy-labeled image segmentation}, in: \bibinfo{booktitle}{International Conference on Medical Image Computing and Computer-Assisted Intervention}, \bibinfo{organization}{Springer}. pp. \bibinfo{pages}{576--584}.

\end{thebibliography}
